\documentclass[aps,prx,reprint,superscriptaddress,amsmath,showpacs,amssymb,longbibliography]{revtex4-1}
\usepackage{graphicx}
\usepackage{amsthm}
\usepackage{amsmath}
\usepackage{amssymb}
\usepackage{dsfont}
\usepackage{bm}
\usepackage{wrapfig}
\usepackage{tikz}
\usepackage{tikz-3dplot}
\usepackage{tkz-euclide}
\usepackage{epstopdf}
\usepackage{pgfplots}
\usepackage{hyperref}
\usepackage{empheq}
\usepackage[T1]{fontenc}
\usepackage{color}
\usepackage{xparse}
\ExplSyntaxOn
\NewDocumentCommand{\mref}{m}{\quinn_mref:n {#1}}
\seq_new:N \l_quinn_mref_seq
\cs_new:Npn \quinn_mref:n #1
 {
  \seq_set_split:Nnn \l_quinn_mref_seq { , } { #1 }
  \seq_pop_right:NN \l_quinn_mref_seq \l_tmpa_tl
  ( 
  \seq_map_inline:Nn \l_quinn_mref_seq
    { \ref{##1},\nobreakspace } 
  \exp_args:NV \ref \l_tmpa_tl 
  ) 
 }
\ExplSyntaxOff

\begin{document}
\newcommand\mpi{\affiliation{Max Planck Institute for Mathematics in the Sciences, Inselstr. 22, D-04103 Leipzig, Germany}}
\newcommand\uleip{\affiliation{Institut f\"ur Theoretische Physik, Universit\"at Leipzig,  Postfach 100 920, D-04009 Leipzig, Germany}}
\newcommand\chau{\affiliation{ 
 Charles University,  
 Faculty of Mathematics and Physics, 
 Department of Macromolecular Physics, 
 V Hole{\v s}ovi{\v c}k{\' a}ch 2, 
 CZ-180~00~Praha, Czech Republic 
}}
\newcommand\lux{\affiliation{Complex Systems and Statistical Mechanics, Department of Physics and Materials Science, University of Luxembourg, L-1511 Luxembourg, Luxembourg}}

\title[]{Active Brownian Heat Engines}
\author{Viktor Holubec}
\email{viktor.holubec@mff.cuni.cz}\uleip\chau
\author{Stefano Steffenoni}\uleip\mpi
\author{Gianmaria Falasco}\uleip\lux
\author{Klaus Kroy}\uleip
\email{klaus.kroy@itp.uni-leipzig.de}

\date{\today} 
\begin{abstract} 
When do non-equilibrium forms of disordered energy qualify as heat? \textcolor{black}{We address this question in the context of cyclically operating heat engines in contact with a non-equilibrium energy reservoir that defies the zeroth law of thermodynamics. To consistently address the latter as a heat bath requires the existence of a precise mapping to an equivalent cycle with an equilibrium bath at a time-dependent effective temperature. We identify the most general setup for which this can generically be ascertained and thoroughly discuss an analytically tractable, experimentally relevant scenario}: a Brownian particle confined in a \textcolor{black}{periodically} modulated harmonic potential and coupled to some non-equilibrium bath of variable activity.  We deduce formal limitations for its thermodynamic performance, including maximum efficiency, efficiency at maximum power, and maximum efficiency at fixed power. They can guide the design of new micro-machines and clarify how much these can outperform passive-bath designs, which has been a debated issue for recent experimental realizations. To illustrate the general principles for practical quasi-static and finite-rate protocols, we further analyze a specific realization of such an active heat engine based on the paradigmatic Active Brownian Particle (ABP) model. This reveals some non-intuitive features of the explicitly computed dynamical effective temperature, 
illustrates various conceptual and practical limitations of the effective-equilibrium mapping, and clarifies the operational relevance of various coarse-grained measures of dissipation.
\end{abstract}

\pacs{05.20.-y, 05.70.Ln} 

\maketitle

\section{Introduction}

The study of heat engines is as old as the industrialization of the
world. Its practical importance has prompted physicists and engineers
to persistently improve their experiments and theories to eventually
establish the consistent theoretical framework of classical thermodynamics.
It allows to quantify very generally, on a phenomenological level, how work is
transformed to heat, and to what extent this process can be
reversed. Heat is the most abundant but least valuable form of
energy, namely ``disordered'' energy dispersed among unresolved degrees of freedom. And turning it into the coherent accessible form called work has been a central aim  since the days of Carnot, Stirling, and other pioneers, after whom some common engine designs have been named.

Recent advances in technology have allowed and also required to extend
this success story into two new major directions.
First, towards microscopic designs that are so small that their
operation becomes stochastic rather than deterministic \cite{san15,
  han09, ozi05, bay15,spe16b}. And secondly to cases where the degrees of
freedom of  the heat bath are themselves driven far from equilibrium,
which potentially matters for small systems operating in a biological
context, e.g., inside living cells or motile bacterial colonies \cite{kri16}.

The analysis of small systems requires an extension of the
theory and basic notions of classical thermodynamics to stochastic
dynamics, which goes under the name of
stochastic thermodynamics \cite{sei05, esp09,sek10, sei12}. It seeks to define heat, work, and entropy on the
level of individual stochastic trajectories. The theory recovers the laws of thermodynamics
for ensemble-averaged quantities but allows to additionally quantify the probability of rare
large fluctuations \cite{sei12}.  Along these lines, many experimental \cite{Abah2012,Rossnagel2016,Brantut2013,Bechinger2012,mart15,Martinez2017} and theoretical \cite{Esposito2010,Verley2014,Dorfman2013,Holubec2017a,
  Campisi2016,Chvosta2010,Kosloff2017,Whitney2014,sch08,hol14} studies have recently been
devoted to microscopic thermodynamic cycles.  In this field, Brownian heat engines play a
paradigmatic role \cite{Martinez2017,Bechinger2012,mart15,sch08,hol14}.
They are usually based on a diffusing colloidal particle that represents the working
substance. Its solvent provides a natural equilibrium heat bath, and a time-dependent confinement
potential can be realized by optical tweezers \cite{Bechinger2012,mart15,hor12}.

\textcolor{black}{Over the last few years, increasing effort has also been devoted to the second mentioned extension of the classical designs, namely to endow quantum~\cite{Rosnagel2014,Niedenzu2018} and
classical (colloidal)~\cite{kri16,zak17,mar18,Kumari2019} heat engines with so-called ``active'' (non-equilibrium)
baths. Some paradigmatic realizations of such
active baths are provided by suspensions of self-propelling bacteria or synthetic microswimmers \cite{Gompper2016,kri16}. Remarkably, they are driven far from equilibrium on the  level of the individual particles --- not merely by
externally imposed overall boundary or body forces.  
The corresponding ``active heat engines'' utilizing such baths can outperform classical designs by evading the zeroth law of thermodynamics, which would require interacting degrees of
freedom to mutually thermalize. 
Engines that exploit this unconventional property can operate between hugely different (effective)
temperatures and thereby at unconventionally high efficiencies,
without risking the evaporation or freezing of the
laboratory. While technically potentially desirable, the lack of thermalization jeopardizes the 
unambiguous distinction of heat from work (roughly speaking, as the contagiously spreading versus the 
coherently preservable form of energy). Which means that one has to resort to the second law of thermodynamics, alone, for that purpose. But active heat engines have even prominently been claimed to transcend the universal performance bounds set by the second law~\cite{kri16}, a notion that is critically examined below.}

\textcolor{black}{
In the following, we first provide a general discussion of heat engines in contact with non-equilibrium reservoirs. The main claims are exemplified by an analytical discussion of a still quite general limiting case: the linear theory for a Brownian heat engine with a non-equilibrium bath.   In particular, in 
Section~\ref{sec:effect_temp} we derive the effective temperature~\eqref{eq:Teffgen} for this class of models, thereby establishing, as a main result, the explicit mapping of the
active-bath engine to a classical engine with an equilibrium bath achieving the same thermodynamic output and performance. Finally, to illustrate and further elucidate our general results and conclusions, Sec.~\ref{sec:example} provides a detailed analysis of
a specific realization of such liner active Brownian heat engine based on the standard minimal model for active particle suspensions, namely the so-called ABP (``active Brownian particle'') model \cite{cat12}. To facilitate the distinction between the general linear theory and the exemplifying model, we refer to the latter by the reminiscent acronym ABE (``active Brownian engine''), in the following. It still allows for several alternative physical interpretations~\cite{Dabelow2019,Crosato2019}, detailed in Sec.~\ref{sec:entropyproduction}. Their dissimilar contributions to the entropy production denounce the non-equilibrium character of the engine that persists during nominally reversibly operation. In Sec.~\ref{sec:ABE-performance}, we analyze the quasi-static and finite-time performance of the model and highlight some peculiarities of the effective temperature.
For better readability, various technical details have been deferred to an Appendix. 
}

\textcolor{black}{
\section{Active heat engines}\label{sec:GEN}
}

\textcolor{black}{
\subsection{Work-to-work versus heat-to-work conversion}
}
\textcolor{black}{
Speaking of non-equilibrium heat baths that defy the zeroth law, an important qualification needs to be made as to how their energy is accessed, if it is no more obliged to spread indiscriminately by itself. Any thermodynamic entity that can qualify as a non-equilibrium bath should be in a non-equilibrium steady state while being able to exchange some disordered form of energy with the so-called system or working medium. Importantly, the exchanged energy should not entirely be work in disguise. In other words, the internal non-equilibrium structure of the bath should not entirely be resolved by the device that feeds on it, in order to allow us to speak of an engine that operates by \emph{heat-to-work conversion}. 
}
\textcolor{black}{
Yet, to exploit the advantages of active baths relative to conventional equilibrium baths, practical designs often rectify at least some of the bath energy by directly tapping some of the internal thermodynamic fluxes that are responsible for the non-equilibrium character of the bath. 
Typical examples are provided by so called \emph{steady-state designs}, such as various flywheels and ratchet-like devices in active suspensions~\cite{Sokolov2010,dil10,nik16,rei16,Vizsnyiczai2017}. They geometrically rectify the persistent motion of active particles and thereby extract work from their (collective) motion against an external load~\cite{Pietzonka2016,Pietzonka2019}.   
Such rectification is reminiscent of the action of a Maxwell demon, but can be less sophisticated, since it feeds on palpable non-equilibrium fluxes rather than feeble equilibrium fluctuations. Yet, it is not immediately obvious whether to classify it as \emph{heat-to-work  conversion} or \emph{work-to-work conversion}. Especially if the rectified nonequilibrium flux in the active bath is driven mechanically or chemically, one is tempted to argue that the rectification should be addressed as a form of work-to-work conversion. However, any heat engine ultimately draws its power from a nonequilibrium thermodynamic flux, namely a heat flux. So, in particular if the rectified flux in the active bath is ultimately caused by a temperature gradient, such as in hot Brownian motion or hot microswimmers~\cite{Rings.2010,fal14,Kroy2016,Geiss2019b}, the notion of heat-to-work conversion in the spirit of two-temperature (Feynman-Smoluchowski) ratchets~\cite{Smoluchowski1927,Feynman2011,rya16,hol17,lee2017,Kalinay2018} also seems very justifiable.
}

\textcolor{black}{In the following, we focus on the operational scheme of traditional heat engines, which cannot extract work from a single bath with time-independent parameters, and are therefore operated cyclically.  We assume that the working medium of the engine is a small (i.e., Brownian) system described by an (overdamped) Hamiltonian $\mathcal H(\mathbf{k},\mathbf{x})$, which depends on a set of stochastic coordinates $\mathbf{x} = (x_1,\dots,x_{N_x})$ and a set of externally controlled parameters $\mathbf{k} = (k_1,\dots,k_{N_k})$, measuring, for example, height of a weight in a gravitational field. These parameters are used to extract work 
(``ordered'' energy in the sense of the external handling) from the engine or to feed it from an external work source. Examples from this class of \emph{cyclic engines} are various colloidal engines immersed in active fluids such as bacteria suspensions (see Ref.~\cite{kri16} for an experiment and Refs.~\cite{zak17,Saha2018,Saha2019,Kumari2019,Ekeh2020} for theoretical works).  We argue that, for these machines, there is a well-defined regime, where energy extracted from the non-equilibrium bath and transformed to work can unambiguously and quantitatively  be interpreted as (a generalized form of) heat --- namely, if there exists a precise mapping to an equivalent setup with an equilibrium bath at a suitable (finite) time-dependent effective temperature $T_\text{eff}(t)$.  Due to the non-equilibrium character of the bath, such engines can still exploit similar ``rectification loopholes'' as the mentioned steady-state ratchets. But the effect is then fully quantified by $T_\text{eff}(t)$, which can, in a precise sense, interpolate between the limits of pure heat-to-work and work-to-work conversion, attained for $T_\text{eff}(t)\equiv T= \text{constant}$ and $\max T_\text{eff}(t)-\min T_\text{eff}(t)\to \infty$, respectively.
}

\textcolor{black}{
\subsection{Energetics and efficiency of cyclic heat engines}\label{sec:energetics}
}

\textcolor{black}{For arbitrary dynamics, the
instantaneous internal energy $\mathcal{H}(t) = \mathcal{H}(\mathbf{k}(t),\mathbf{x}(t))$ of the working medium of the engine changes as
\begin{equation}
\frac{d}{dt} \mathcal{H}(t)
= \dot{w}(t) + \dot{q}(t)
\label{eq:dH}
\end{equation}
with
\begin{eqnarray}
\dot{w}(t) &=& \sum_{i=1}^{N_k} \frac{\partial}{\partial k_i}\mathcal{H}(t) \dot{k}_i(t),
\label{eq:wH}\\
\dot{q}(t) &=& \sum_{i=1}^{N_x} \frac{\partial}{\partial x_i}\mathcal{H}(t) \dot{x}_i(t) = \frac{d}{dt} \mathcal{H}(t) - \dot{w}(t).
\label{eq:qH}
\end{eqnarray}
The contribution $\dot{w}$ corresponds to a change of the externally controlled parameters 
$\mathbf{k} = \mathbf{k}(t)$ and thus it is naturally identified as work delivered to the working medium from the external work reservoir~\cite{sch08,sek10,sei12,Bechinger2012,hol14,mart15,kri16,zha17,Ekeh2020}. The remaining part of the energy change, $q$, is then acquired from the heat reservoirs. In accord with the standard heat engine nomenclature, it is identified as heat --- with the above-mentioned potential caveats in mind.
}

\textcolor{black}{
The above defined work and heat transfers are stochastic quantities that fluctuate due to the stochastic nature of the coordinates $\mathbf{x}$. One is often interested in their mean values both over a certain span of time and over the stochastic ensemble. Upon integration over time and ensemble averaging, the average total work exchanged between the system and its environment during the time interval $(t_{\rm i}, t_{\rm f})$ is given by
\begin{equation}\label{eq:work}
W(t_{\rm i},t_{\rm f}) =  \int_{t_{\rm i}}^{t_{\rm f}}dt\, \dot{W}(t) = \int_{t_{\rm i}}^{t_{\rm f}}dt\, \left<\dot{w}(t)\right>
\end{equation} 
and the corresponding total heat by
\begin{equation}\label{eq:heat}
Q(t_{\rm i},t_{\rm f}) = \int_{t_{\rm i}}^{t_{\rm f}}dt\, \dot{Q}(t) = \int_{t_{\rm i}}^{t_{\rm f}}dt\, \left<\dot{q}(t)\right>
\end{equation}
}

\textcolor{black}{
From now on, we assume that the parameters of the Hamiltonian are varied periodically, with period $t_{\rm p}$. The (ensemble-averaged) states of the system and the reservoirs are assumed to eventually attain a time-periodic limit cycle with the same period. If not explicitly written otherwise, all variables below will be evaluated on this limit cycle.}

\begin{figure}[tb]
\centering
\includegraphics[width=1.0\columnwidth]{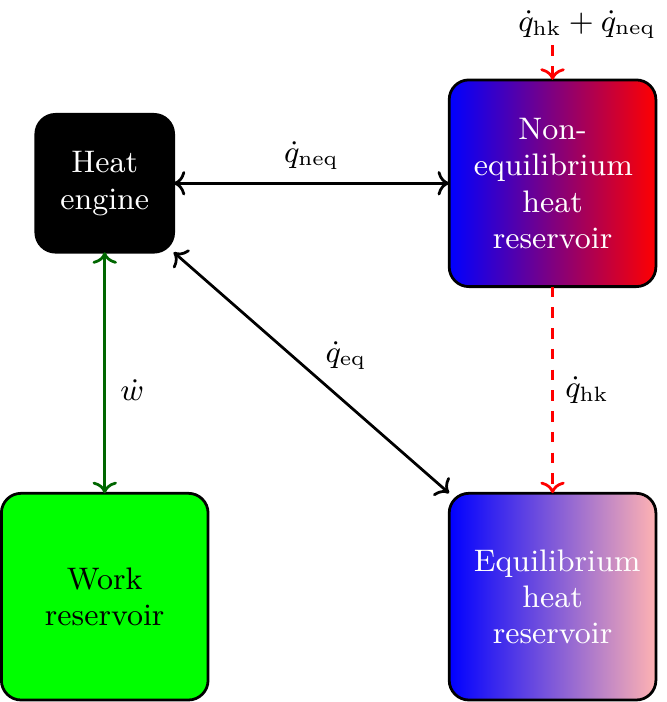}
\caption{\textcolor{black}{Cyclic heat engine transforming the heat flux $\dot q = \dot q_{\rm neq} + \dot q_{\rm eq}$ from a non-equilibrium (neq) and equilibrium (eq) heat reservoir into usable power $\dot w$. The corresponding energy fluxes relevant for the engine's operation are depicted by arrows.  The dashed arrow depicts the housekeeping heat flux, $\dot{q}_{\rm hk}$, flowing from the active bath to the infinite equilibrium reservoir, which prevents the active bath from overheating. This energy flux and also $\dot{q}_{\rm neq}$ are sustained by the energy influx $\dot{q}_{\rm hk} + \dot{q}_{\rm neq}$ into the non-equilibrium bath, which keeps it in a non-equilibrium steady state. In this paper, we discuss the setup where the energy $\mathcal{H}[\mathbf{k}(t)]$ of the working medium of the engine (e.g. a single trapped colloid) is periodically modulated by an external control parameter $\mathbf{k}(t)$ and the temperature/activity (technically: the noise intensities) of the two heat reservoirs.}
  }
\label{fig:HEscatch}
\end{figure}

\textcolor{black}{
The net average work performed or \emph{output work} by the engine per cycle is, with the above definitions, expressed as 
\begin{equation}
W_{\rm out} = -W(0,t_{\rm p}).
\label{eq:WH}
\end{equation}
As the \emph{input heat}, $Q_{\rm in}$, one usually identifies only the heat acquired during those parts of the cycle during which heat on average flows from the bath into the system~\cite{Callen2006}, i.e. when $\Theta(\left<\dot{q}\right>) > 0$, where $\Theta$ denotes the Heaviside step function. So we have
\begin{equation}
Q_{\rm in} = \int_0^{t_{\rm p}} dt\, \dot{Q}(t)\Theta[\dot{Q}(t)]\,,
\label{eq:qin}
\end{equation}
which may well differ from $Q(0,t_{\rm p})$.
So while the definition may look a bit awkward, it allows us to write the input heat in a form that is independent of specific details of the driving protocol. For standard thermodynamic cycles such as Carnot or Stirling cycle with a hot and a cold equilibrium heat bath, it recovers the standard expressions for the heat afforded via the hot reservoir (irrespective of the amount of heat taken up by the cold reservoir).
}

\textcolor{black}{
Common measures of performance of a heat engine are its output power $P$ and efficiency $\eta$: 
\begin{align}
&P \equiv \frac{W_{\text{out}}}{t_{\rm p}},\,&& \eta \equiv \frac{W_{\text{out}}}{Q_{\text{in}}}.
\label{eq:PandEta}
\end{align}
If the heat bath consists of a non-equilibrium reservoir, one should in principle add the housekeeping contribution $\int_0^{t_{\rm p}} dt\, \dot{q}_{\rm hk}$ that maintains its non-equilibrium steady state, in the denominator (cf.\ Fig.~\ref{fig:HEscatch}). In most practical settings, this contribution would completely overshadow $Q_{\rm in}$ --- rendering the efficiency tiny and potentially dependent on the technical realization of reservoir, which is usually not desirable. In accord with earlier works~\cite{kri16,zak17}, we therefore use the definition in~\eqref{eq:PandEta} also in this case and identify as heat only the energy actually exchanged with the reservoirs ($\dot q$ in Fig.~\ref{fig:HEscatch}), rather than the whole energy afforded to operate the engine and its active bath  ($\dot{q}_{\rm hk} + \dot{q}_{\rm neq}$ in Fig.~\ref{fig:HEscatch}).
}

\textcolor{black}{
If the engine instead communicates with an equilibrium bath at temperature $T(t)$, its efficiency is unambiguously restricted by the first and second law of thermodynamics to obey $\eta \le 1$ and $\eta \le \eta_{\infty} < \eta_{\rm C}  = 1 - \min{(T)}/\max{(T)} \le 1$, respectively. Here, $\eta_{\infty}$ refers to the value obtained upon infinitely slow, reversible operation, and $\eta_{\rm C}$ is the Carnot efficiency. On the level of stochastic heat and work transfers, these constraints are moreover reflected by various fluctuation theorems for the corresponding probability distributions~\cite{jar97,jar97b,cro98,cro99,sei12}.}

\textcolor{black}{
Allowing for an (additional) active bath, the interpretation of the conventional formalism may require some extra considerations. First, one can exploit the non-equilibrium state of the bath to effectively isolate certain degrees of freedom from the rest of the setup, thereby effectively circumventing the zeroth law. This allows one to emulate unusually high or low temperatures (for these degrees of freedom) without contaminating many others, and thus to reach exceptionally high efficiencies.
An example would be a hot Brownian swimmer, which is actually laser heated relative to the solvent by only a few Kelvin, while executing a random motion as if it had been heated by thousands of Kelvin, which would technically be much more difficult to achieve for a conventional equilibrium bath obeying the zeroth law \cite{Geiss2019b}. }

\textcolor{black}{
Secondly, one can extract net work from a single steady-state heat bath at constant activity, thus apparently beating the second law~\cite{Ekeh2020}.
For this one needs at least two control parameters, though, since quasi-statically operating engines with a single control parameter $\mathbf k = k$ allow the output power to be integrated  $\left<\dot{w}(t) dt\right> = \left<\partial\mathcal{H}/\partial k \right>dk = f(k)dk$. Here, $f(k)$ depends solely on $k$ since all other parameters are held constant. A physically sensible one-dimensional function $f(k)$ can always be written as a derivative $f(k) \equiv  \text{d} g(k)/\text{d}k$. The output work per cycle then reads: $W_{\rm out} = \int_{0}^{t_{\rm p}}\text{d}t\, \langle\dot{w}(t)\rangle  =  g(k(t_{\rm p}))-g(k(0)) = 0$, because of the periodicity of $k(t)$. This result is valid regardless of the properties of the steady-state bath, except that for non-quasi-static protocols the output work will be negative, due to finite-time losses. 
For two (and more) parameters, on the other hand, $\left<\dot{w}(t)\right>dt = \sum_{i=1}^{N_{k}}f_i[\mathbf{k}(t)]\text{d}k_i$. Hence $W_{\rm out} = 0$ now only holds if an integrability condition is satisfied, namely that a function $g[\mathbf{k}(t)]$ exists such that $f_i = \partial g/\partial k_i$ is a gradient and thus $\left<\dot{w}(t)\right>\text{d}t= \nabla g \cdot d\mathbf{k}$.  Otherwise, internal currents may indeed allow the extraction of work from a single non-equilibrium bath at constant activity~\cite{Ekeh2020}. 
}

\textcolor{black}{
In both of the above examples of how to ``beat'' classical constraints on the performance of heat engines an equivalent of temperature is seen to play a crucial role, namely the one characterizing the Brownian motion of the microswimmer and the one characterizing the constant activity of the active bath, respectively. Indeed, as we lay out in the following paragraph and in even greater detail in the remainder of this contribution, this notion can sometimes be made fully quantitative and then be used to explicitly compute meaningful efficiencies for active heat engines. 
}

\textcolor{black}{
\subsection{Dynamic effective temperature}\label{sec:GENEffT}
}

\textcolor{black}{
The crucial step is to construct a mapping for the power $P$ and efficiency $\eta$ of a heat engine in contact with a non-equilibrium bath to that of a heat engine in contact with an equilibrium bath. This can be achieved if one can define a temperature in the sense of the second law of thermodynamics~\cite{casaz.2003,Geiss2019b}. Which is the case if, for a given protocol for varying the control parameters $\bf k(t)$, the energy fluxes $\left<\dot{w}(t)\right>$ and $\left<\dot{q}(t)\right>$ for the heat engine in contact with the non-equilibrium bath agree with those for a virtual heat engine in contact with an equivalent equilibrium bath maintained at a time-dependent temperature $T_{\rm eff}(t)$. This then allows the application of known results for heat engines with equilibrium baths to meaningfully define and assess the performance of active engines. Per construction, their efficiencies are then bounded by the second law. (Further consequences of the mapping are discussed for a specific example in Sec.~\ref{sec:effect_temp}.)
}

\textcolor{black}{The most general situation for which an appropriate effective temperature can always be found is when one can write the Hamiltonian in the form $\mathcal{H} = k(t) h(\mathbf{x})$, with an arbitrary function $h(\mathbf{x})$ diverging at $|x| \to \infty$. Then we have $\left<\dot{w}\right> = \dot{k}(t) f(t) $ and $\left<\dot{q}\right> = k(t) \dot{f}(t)$, with $f(t) = \left<h(\mathbf{x}) \right>$. In general, for a non-equilibrium bath described by a set of functions ${\mathbf b}(t)$, $f(t)$ is a functional of the external protocol $k(t)$ and the bath parameters ${\mathbf b}(t)$, say $f(t)  = f_{\rm neq}[\{k(t),{\mathbf b}(t)\}_{t=0}^{t_{\rm p}}]$.
If the equilibrium mapping exists, this functional can be written as $f(t)  = f_{\rm eq}[\{k(t),T_{\rm eff}(t),\gamma(t)\}_{t=0}^{t_{\rm p}}]$, where all relevant effects of the bath parameters have been subsumed into the dynamic effective temperature $T_{\rm eff}(t)$ and possibly also a time-dependent friction $\gamma(t)$.} 

\textcolor{black}{
These two quantities are implicitly given by the functional identity $f_{\rm eq}(t) \equiv f_{\rm neq}(t)$, which has to be solved to derive their explicit form. In the next section, we discuss a specific scenario where this can be always achieved analytically. In general, our physical intuition suggests that the equation $f_{\rm eq}(t) \equiv f_{\rm neq}(t)$ has at least one solution. While it may be difficult to rigorously prove its existence and uniqueness on such a general level, what matters most with respect to the thermodynamic performance is the case of quasistatic driving. In this limit, $f_{\rm eq}(t)$ is only a function of $k(t)$ and $T_{\rm eff}(t)$, which can be determined by calculating the average $f_{\rm eq} = \left< h({\bf x})\right>$ over the Gibbs canonical distribution $p({\bf x},t) = \exp[- k(t) h({\bf x})/T_{\rm eff}(t)]/Z$, where $Z$ is the normalization constant, and 
we have set the Boltzmann constant to unity, $k_{\rm B} \to 1$, measuring
energies in Kelvin. The resulting equation $f_{\rm eq}(t) \equiv f_{\rm neq}(t)$ for $T_{\rm eff}(t)$ can then always be solved, because any value of the average $\left< h(\bf{x})\right>$ (taken over the Gibbs distribution) can be assigned an effective temperature $T_{\rm eff}(t)$ varying between zero and infinity, thereby exhausting all possible values of the average obtainable with an arbitrary non-equilibrium distribution.}

\textcolor{black}{For more general Hamiltonians of the form $\mathcal{H} = k(t) h_1(\mathbf{x}) +
h_2(\mathbf{x})$, or even more complicated,  we have $\left<\dot{w}\right> = \dot{k} f_1 $ and $\left<\dot{q}\right> = k \dot{f}_1 + \dot{f}_2$, with $f_i = \left<h_i(\mathbf{x}) \right>$ again being functionals of the driving and bath parameters. These two functionals need not consistently determine a single function $T_{\rm eff}$,  solving $f_{i, \rm eq} \equiv f_{i, \rm neq}$ for both $i = 1,2$. Then, equivalent cycles with equilibrium baths might still exist under specific circumstances, but they are not generally guaranteed or generically expected, anymore.}

\textcolor{black}{To sum up this introductory section and to answer the question asked in the beginning of our abstract, we note that the notion of heat can unambiguously be generalized to non-equilibrium situations where the zeroth law does not hold, but it is tied to an operational definition of an effective temperature in the sense of the second law. In other words, one has to require that the only energy that can be extracted from a non-equilibrium heat bath is of the disordered form that comes with a reduced work efficiency. Otherwise, one actually deals with some sort of work reservoir in disguise. Active heat engines coupled to such non-equilibrium baths and Hamiltonians proportional to a single control parameter can always be reinterpreted in terms of equivalent engines in contact with equilibrium baths, at some dynamic effective temperature. Their thermodynamic properties thus obey standard-second law bounds, with important consequences for the interpretation of experimental results.
} 

\textcolor{black}{
\subsection{Application to experimental data}
\label{sec:experiment}
}

\textcolor{black}{
A relevant real-world realization of a heat engine in contact with a non-equilirbium bath is the bacterial heat engine of
Ref.~\cite{kri16}. In this impressive experimental study, a colloidal
particle with Cartesian position $\{x,y\}$ was trapped in a time-dependent harmonic potential, 
\begin{equation}
\mathcal{H}(x,y,t) = \frac{1}{2}k(t) \mathbf{r}^2 =  \frac{1}{2}k(t)(x^2+y^2),
\label{eq:potential}
\end{equation}
and immersed in a bath of self-propelled bacteria. Both the trap stiffness $k(t)$ and the bacterial activity were quasi-statically modulated to realize a Stirling-type active heat engine with a cycle composed of
two isochoric and two isothermal state changes. These were technically implemented by changing the bacterial activity at constant trap stiffness $k$ and vice versa, respectively.} \textcolor{black}{The ensuing colloid dynamics was observed to converge
to a quasi-static limit cycle transforming energy absorbed from the disordered bacterial bath into colloidal work.} 

The authors measured
the work done per cycle as well as the energy (heat) obtained per
cycle from the bath and determined the efficiency of the machine as their ratio. 
\textcolor{black}{
From Eq.~(\ref{eq:potential}) the time-dependent average system energy reads
\begin{equation}
\left<\mathcal H\right> = \frac{1}{2}k(t)\left[\sigma_x(t) + \sigma_y(t)\right] = \frac{1}{2}k(t)\sigma(t),
\label{eq:internal_energy}
\end{equation}
where $\sigma_x = \left<x^2\right>$, $\sigma_y = \left<y^2\right>$,
and $\sigma = \left<\mathbf{r}\cdot\mathbf{r}\right>$. Due to the
symmetry of the potential, the average particle displacements
$\left<x\right>$ and $\left<y\right>$ vanish during the cycle,
so that the mean square displacements $\sigma_x(t)$ and $\sigma_y(t)$ 
also determine the long-time variances for the $x$- and $y$-coordinates, respectively.}

\textcolor{black}{
Based on their analysis of the apparent equipartition temperature $T_{\rm eff}(t) \equiv k(t)\sigma (t)/2$, denoted by $T_{\rm a}$ in Ref.~\cite{kri16}, its authors concluded that they had realized a Stirling cycle that allowed them to significantly surpass the maximum Stirling efficiency, $(1+1/\log (k_>/k_<))$ attained for equilibrium heat baths with an infinite temperature difference ($k_>$ and $k_<$ denote maximum and minimum values of $k(t)$ during the cycle). This extraordinary result was attributed to large non-Gaussian fluctuations in the non-equilibrium bacterial reservoir, which, according to the authors, cannot be captured by an effective temperature.}

\textcolor{black}{These conclusions are plainly at odds with the general analysis in the preceding paragraph. To see this, notice that the experimental heat engine corresponds to a Hamiltonian proportional to a single control parameter, for which one can always define an effective temperature so that the conventional bounds on the efficiency apply. Using Eqs.~\eqref{eq:wH} and \eqref{eq:qH} (employed also in Ref.~\cite{kri16} to evaluate work and heat fluxes into the system), we obtain $\left<\dot{w}(t)\right> = \dot{k}(t) \sigma(t)/2$ and $\left<\dot{q}(t)\right> = k(t) \dot{\sigma}(t)/2$.  The  equivalent heat engine with an equilibrium bath has the bath temperature $T_{\rm eff}(t)$. It thus has the same energy input (heat), as correctly noted in Ref.~\cite{kri16}, but also the same energy output (work).
Accordingly, if $T_{\rm eff}(t)$ evolves along a Stirling cycle, the efficiency $\eta$ of the active engine, determined by the ratio of output work over afforded heat, is necessarily bounded by the Stirling efficiency. The non-Gaussian fluctuations in the bath indeed affect the output work, input heat, and efficiency of the engine, but only via the mean square displacement $\sigma$, hence again via the appropriate effective temperature $T_{\rm eff}$. Assuming that heat and work were accurately measured (which is supported by the correctly measured Stirling efficiency in the case of inactive bacteria), the observation of an efficiency surpassing the maximum value for Stirling engines calls into question the notion that the experimental engine realized a Stirling cycle with respect to $T_{\rm eff}$ (see also Ref.~\cite{zak17}). As we demonstrate next, the dynamic effective temperature $T_{\rm eff}(t)$ may generally indeed vary in time even while the ambient solvent temperature and the activity remain constant. }

\section{Linear theory: Dynamics}


\begin{figure}[tb]
\centering
\begin{tabular}{ll}
\includegraphics[scale=0.64]{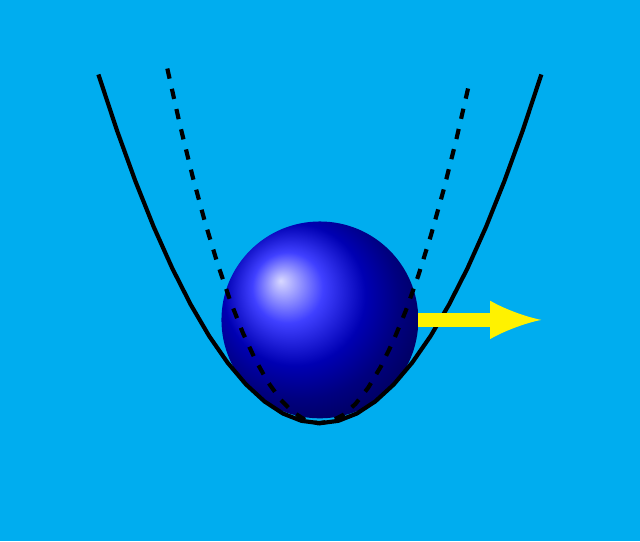}
&
\includegraphics[scale=0.64]{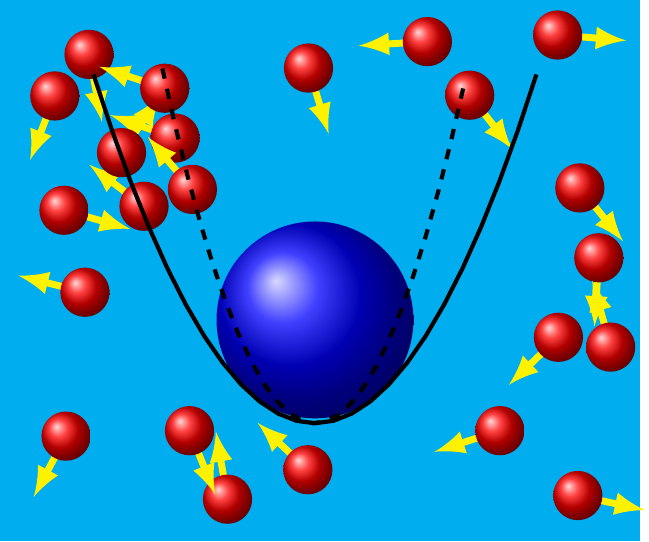}
\end{tabular}
\caption{Schematic designs of microscopic heat engines based on
  colloids in modulated harmonic traps, playing the roles of the
  working substance and the movable piston, respectively. 
  Left: active particle in a ``passive'' equilibrium bath. Right:
  passive particle in an ``active''  non-equilibrium bath
  composed of energy consuming micro-swimmers immersed into a passive
  background fluid. To operate the heat engine, the bath temperature
  and/or activity as well as the confinement strength are modulated
  cyclically. Thereby ``disordered'' energy dispersed in the bath and randomly propelling the colloid against its confinement is concentrated in a degree of freedom that can be externally harnessed to perform (mechanical) work.}
\label{fig:HE}
\end{figure}

\textcolor{black}{
Up to this point, we have not specified any particular system dynamics and thus the described results are valid for arbitrary time-evolution of the degrees of freedom $\mathbf x$. To provide better insight and to show that a non-intuitive behavior of effective temperatures can be expected, this section investigates a specific (but from the point of view of Brownian heat engines still quite generic) exactly solvable class of models. Concretely, we analytically derive the effective temperature for a class of one-parameter engines inspired by the experimental work described above. We detail the mapping to the equilibrium model and its consequences for the thermodynamics of the active heat engine. In particular, we reveal a nontrivial behavior of the effective temperature. This seems to be the first explicit result of its type.}

From now on, we specialize our discussion to a heat engine consisting of a colloidal particle confined to
a \emph{time-dependent harmonic} potential with an externally controlled stiffness $k(t)$, as introduced in Eq.~(\ref{eq:potential}).  We specify the dynamics
by further requiring that the colloid is immersed in
a (possibly) non-equilibrium bath, which couples to it 
via a drag coefficient $\mu^{-1}$ and a zero-mean
\emph{additive} noise $\bm{\eta}(t)$, so that its position $\mathbf{r}=(x,y)^{\!\top}$ obeys the \emph{overdamped
  linear} Langevin equation  
\begin{equation}
\dot{\textbf{r}} = - \mu k(t) \textbf{r}(t) + \bm{\eta}(t).
\label{eq:linear_model}
\end{equation}
Depending on the noise correlations, which remain to be prescribed and need not be Markovian,
this equation can describe various experimentally
relevant situations. In Fig.~\ref{fig:HE} we depict two of them that
we discuss further below: namely, an 
active particle or ``microswimmer'' immersed in a passive equilibrium 
bath (left) \cite{fil12,sza14}, and a (passive) colloid
immersed in an active non-equilibrium bath that is itself composed of active particles
swimming in a thermal background solvent
(right) \cite{wu00,zha17,ang11,har14,far15}. Further examples are provided by devices that share the same
formal description on a suitably coarse-grained level, such as noisy
electric circuits and similar Langevin systems~\cite{Coffey.2004}.

In line with such realizations, the trapping potential \eqref{eq:potential}
has the harmonic standard form experimentally created with the help
of optical tweezers \cite{Bechinger2012,mart15,kri16}. We have also taken
advantage of the fact that such experiments are typically designed
in a quasi two-dimensional geometry, in narrow gaps between two glass coverslips.
For simplicity, the particle mobility is represented by a constant scalar
 $\mu$  and the two-time correlation matrix 
\begin{equation}\label{eq:noise_covariance}
C_{ij}(t,t')\equiv
\left<\eta_i(t)\eta_j(t')\right> \propto \delta_{ij}    
\end{equation}
of the noise $\bm{\eta} = (\eta_x, \eta_y)$ by a diagonal form.
Our analysis can of course straightforwardly be generalized to
arbitrary dimensions and mobility matrices. 

If $\bm \eta$ in Eq.~\eqref{eq:linear_model} stands for the white noise, the model provides a good description for existing experimental realisations of Brownian heat engines~\cite{Bechinger2012,mart15}. Their thermodynamics has been thoroughly analyzed in the literature~\cite{sch08,Holubec2017,Holubec2018}.
An example for an experimental realisation of the non-equilibrium-noise version is the active Brownian engine with a bacterial bath~\cite{kri16} discussed in the previous section. The performance of a quasi-static Stirling heat engine based on the latter design was already nicely analyzed by Zakine et al.~\cite{zak17}. Its finite-time performance was numerically investigated in Refs.~\cite{Saha2018,Saha2019,Kumari2019}. With respect to these studies, which employ specific protocols, our approach is valid for arbitrary driving protocols at arbitrary speeds.

As a main result, we show in the following that the thermodynamics of
the system described by Eq.~\eqref{eq:linear_model} with a non-equilibrium noise $\bm{\eta}$,
to which we refer as the (linear) \emph{active} heat engine, can be mapped onto the
well-investigated model with a passive equilibrium bath
\cite{sch08,Holubec2017,Holubec2018}, to which
we refer as the \emph{passive} model:
\begin{equation}
\dot{\textbf{r}}(t) = - \mu k(t) \textbf{r}(t) + \sqrt{2 D_{\rm eff}(t)}\bm{\xi}(t).
\label{eq:eq_model}
\end{equation}
Its bath is characterized by the Gaussian white noise
$\bm{\xi}(t)$ with zero mean, $\left<\bm{\xi}(t)\right> = 0$, the unit
correlation matrix matrix, $\left<\xi_i(t) \xi_j(t')
\right>=\delta_{ij}\delta(t-t')$, and a time-dependent (effective) temperature~\footnote{Here and in the rest of the paper, we use the Stratonovich convention. See Sec.~\ref{eq:eff_temp_ABP} for an explicit calculation of the effective temperature for the ABE model.}
\begin{equation}
T_{\rm eff}(t) = \frac{D_{\rm eff}(t)}{\mu} = \frac{1}{2\mu}\left<\textbf{r}(t)\cdot \bm{\eta}(t) \right>.
\label{eq:Teffgen}
\end{equation}
Below, the latter is shown to follow solely from the two-time
correlation matrix $C(t,t')$ of the noise $\bm{\eta} = (\eta_x,
\eta_y)$. Since the passive model~\eqref{eq:eq_model} and the
corresponding temperature~\eqref{eq:Teffgen} describe the active model
only \rm{effectively}, in terms of its average thermodynamic
properties, \eqref{eq:eq_model} and \eqref{eq:Teffgen} are referred
to as an effective passive model and an effective temperature, respectively.  

The existence of this mapping immediately implies that the performance
of the active heat engine in terms of its output power and efficiency
is precisely that of the corresponding effective equilibrium
model. Therefore, the known bounds on (finite-time) performance of
cyclic Brownian heat engines described by Eq.~\eqref{eq:eq_model}, such as
the ultimate Carnot efficiency bound \cite{callen1985}, the
efficiency at maximum power \cite{sch08}, the maximum efficiency at
arbitrary power \cite{Holubec2016,Holubec2017}, and the possibility to
almost attain the reversible efficiency at nonzero power
\cite{Holubec2018}, directly carry over to the active heat
engine. Furthermore, the effective equilibrium model also sets bounds
on average thermodynamic variables for non-cyclic and even transient
processes. Yet, the non-equilibrium character of the underlying  
dynamics reveals itself upon closer inspection,
as detailed in the remainder of the paper.

\section{Linear theory: effective temperature}
\label{sec:effect_temp}

\textcolor{black}{
\subsection{General initial conditions}
}

\textcolor{black}{
It is a noteworthy property of the linear theory and the experiments that motivate it that thermodynamic quantities like  work, heat, and efficiency are all determined solely by the variance $\sigma(t)$ of the colloidal position, see Sec.~\ref{sec:experiment}. 
The variance $\sigma(t)$ itself obeys the ordinary differential equation }
\begin{equation}
\dot{\sigma}(t) = - 2 \mu k(t) \sigma(t) +  2\left<\mathbf{r}(t)\cdot\bm{\eta}(t)\right>\,
\label{eq:var_formal}
\end{equation}
which follows from Eq.~\eqref{eq:linear_model} by taking the
scalar product with $\mathbf{r}$ on both sides and averaging over the
noise. For arbitrary additive noise $\bm{\eta}$,
Eq.~\eqref{eq:linear_model} has the formal solution
\begin{equation}
\mathbf{r}(t) = \mathbf{r}_0{\rm e}^{-K(t,t_0)} + \int_{t_0}^t dt'\,\bm{\eta}(t') {\rm e}^{-K(t,t')},
\label{eq:r_formal}
\end{equation}
with $K(t,t') \equiv \mu \int_{t'}^t dt''\, k(t'')$ and $\mathbf{r}_0
\equiv \mathbf{r}(0)$ denoting an arbitrary initial position of the
particle. With the two-time noise correlation matrix,
$C(t,t')$ from Eq.~\eqref{eq:noise_covariance}, the average in Eq.~\eqref{eq:var_formal} evaluates to 
\begin{multline}
\left<\mathbf{r}(t)\cdot\bm{\eta}(t)\right> = 2 D_{\rm eff}(t)\equiv 
\left<\mathbf{r}_0\cdot\bm{\eta}(t)\right>{\rm e}^{-K(t,t_0)}
+\\  \int_{t_0}^t dt'\,{\rm Tr}[C(t,t')] {\rm e}^{-K(t,t')} \,,
\label{eq:r_formal2}
\end{multline}
where ${\rm Tr}$ denotes the trace operation. A crucial observation is
that Eq.~\eqref{eq:var_formal} therefore assumes a form that would also result from the Gaussian white noise $\bm{\eta} = \sqrt{2 D_{\rm eff}(t)}
\bm{\xi}(t)$ with the correlation matrix $C_{ij}(t,t') = 2 \sqrt{D_{\rm
    eff}(t)D_{\rm eff}(t')}\delta_{ij}\delta(t-t')$~\footnote{\textcolor{black}{This can be seen by substituting this expression for matrix $C$ into the right-hand side of Eq.~\eqref{eq:r_formal2} and evaluating the integral therein. We further assume that the initial condition is not correlated with the equilibrium noise, $\left<\mathbf{r}_0\cdot\bm{\xi} \right> = 0$, which is quite natural.}}.
This implies that the average thermodynamic behavior of the active
model~\eqref{eq:linear_model} with arbitrary additive noise 
is the same as that of the passive model~\eqref{eq:eq_model} 
with an effective equilibrium bath temperature
\begin{equation}
T_{\rm eff}(t) = \frac{D_{\rm eff}(t)}{\mu} = \frac{\left<\textbf{r}(t)\cdot \bm{\eta}(t) \right>}{2\mu} = \frac{k(t) \sigma(t)}{2} +\frac{\dot{\sigma}(t)}{4\mu}.
\label{eq:Teffgen2}
\end{equation}
The last expression follows from Eq.~\eqref{eq:var_formal}. It shows that also the effective temperature is uniquely given by the variance $\sigma$.
Notably, the result~\eqref{eq:Teffgen2} is valid arbitrarily far from equilibrium and it does not
follow from any close-to-equilibrium linear-response approximation
like in the Green-Kubo formula \cite{kub66}.

Also note that for positive effective temperature $T_{\rm eff}(t) \ge 0$,
Eq.~\eqref{eq:Teffgen2} establishes the announced mapping between the
active and passive heat engine and thus proves our main result.
Negative effective temperatures can however be obtained, for example,
during transients departing from initial conditions with
$\left<\mathbf{r}_0\cdot \bm{\eta}(t)\right> < 0$. At late times,
the sign of the effective temperature is determined by the integral in
Eq.~\eqref{eq:r_formal2}, which is positive for standard correlation
matrices $C(t,t')$ with non-negative diagonal elements. For a
quasi-static process, where the system parameters vary slowly compared
to the intrinsic relaxation times, one can neglect $\dot{\sigma}(t)$ relative to the
other term in Eq.~\eqref{eq:Teffgen2}. The effective
temperature then reduces to the well-known form
\cite{kri16}
\begin{equation}
T_{\rm eff}(t) = k(t)\sigma(t)/2.
\label{eq:qs_Teff}
\end{equation}
For slowly driven systems, the effective temperature is thus always
positive, thanks to the positivity of the trap stiffness $k$ and variance $\sigma$.

\subsection{Cyclic heat engines}

The definition~\eqref{eq:Teffgen2} of the effective
temperature applies both under transient and stationary conditions.
Cyclic heat engines operate time-periodically by virtue of
their periodic driving. Accordingly, we assume that the potential stiffness
$k(t)$ is a periodic function with period $t_{\rm p}$ and that the 
noise correlation matrix is of the form 
\begin{equation}
C_{ij}(t,t') = 2 \delta_{ij}I(t)I(t')f_{i}(t-t'),
\end{equation}
where $I(t)$ stands for a $t_{\rm p}$-periodic intensity of the noise,
and $f_{i}(t)$ are arbitrary functions obeying $f_{i}(0) = 1$ and
decaying towards zero as $t\to\infty$. The system dynamics then
settles onto a time-periodic attractor, independent of the initial
condition $\mathbf{r}_0$, at late times. From now on, we assume that
the engine operates in this ``steady state'' regime, to which we refer
as the \emph{limit cycle}. During the cycle, the effective temperature
$T_{\rm eff}(t)$ takes the form [see Eqs.~\eqref{eq:r_formal2}
 and \eqref{eq:Teffgen2}]
\begin{equation}
\frac{1}{\mu}I(t)\int_{-\infty}^t dt'\,I(t')[f_{x}(t-t')+f_{y}(t-t')] {\rm e}^{-K(t,t')}.
\label{eq:Teffgen3}
\end{equation} 
Importantly, for positive diagonal elements of the correlation matrix,
the effective temperature is then manifestly positive, as required to
map the active onto the passive model.

\subsection{(Im)possible generalizations}

The simplifying power of the present approach crucially relies on two main features. Firstly, on the linearity of Eq.~\eqref{eq:linear_model}, and secondly on the fact that thermodynamics is predominantly concerned with average energetics. 

For the active heat engines discussed in the present contribution, the
pertinent microscopic degree of freedom is the position of the
colloid. Its thermodynamics is contained in the variance $\sigma =
\left<\textbf{r}\cdot\textbf{r}\right>$, which controls the complete
average energetics (work and heat) of the engine through
Eqs.~\eqref{eq:work} and \eqref{eq:heat}. However, the described mapping to a passive-bath model cannot be
extended beyond such average energetics, since the active~\eqref{eq:linear_model} and
passive~\eqref{eq:eq_model} heat engines differ in variables which
depend on higher moments of the position $\mathbf{r}$
or its complete distribution. This is for example the case for 
the total entropy or the fluctuations of work, heat and entropy. Without further ado, one thus cannot take for granted the results
obtained under the assumption of a perfect contact with an equilibrium
bath, such as the Jarzynski equality \cite{jar97}, the Crooks fluctuation theorem
\cite{cro99}, the Hatano-Sasa equality \cite{hat01,Szamel2019}, and various 
inequalities containing higher moments of work, heat, and entropy, such
as thermodynamic uncertainty relations \cite{Gingrich2016,
  Proesmans2017, Pietzonka2017, Falasco2019}.

Also note that, for a true equilibrium noise $\bm{\eta}$, 
the (effective) temperature $T_\text{eff}$ in Eq.~\eqref{eq:Teffgen} would agree with all other possible
definitions of temperature, thereby tying together many \emph{a priori} unrelated dynamical quantities (e.g.\ by their structurally identical Boltzmann distributions or fluctuation-dissipation theorems, etc.). However, for a non-equilibrium noise, differently defined temperatures can (and generally will) have different values. We refer to Refs.~\cite{Cugliandolo1997,fal14,falasco2014non,wul17,Marconi2017,Chaki2018,Caprini2019,Chaki2019,Caprini2019b} for various (complementary) approaches to effective temperatures and Refs.~\cite{casaz.2003,cug11,Geiss2019b} for some reviews.
Moreover, as illustrated by the ABP results 
\eqref{eq:Langevin1A}--\eqref{eq:Langevin1B} in App.~\ref{appx:PDF}, typical non-equilibrium distributions deviate
strongly from  Boltzmann's Gaussian equilibrium distribution,
such as the one characterizing the long-time limit of the equilibrium process, Eq.~\eqref{eq:eq_model}, at constant $T_{\rm eff}$ --- namely
$\rho(\textbf{r}) \propto \exp[- k\textbf{r}^2/2T_{\rm eff}]$. 
Therefore, in order to build an
effective thermodynamic description from a non-equilibrium
statistical-mechanics model, one generally has to
calculate precisely the effective temperatures corresponding to the
relevant degrees of freedom, under the prescribed conditions.

This leads to the mentioned second limitation of the presented effective-temperature mapping, namely that it hinges on the linearity of the model. To make the point, let us consider a
one-dimensional setting with the potential $\mathcal{H}(x,t) = k(t)x^n/n$ when
the Langevin equation for position $x$ reads
\begin{equation}
\dot{x}(t) = - k(t) [x(t)]^{n-1} + \eta(t)
\label{eq:Langevin_nonlinear}
\end{equation}
and the internal energy, work, and heat (per unit time) are given
by $\left< \mathcal{H} (t)\right> = k(t)\left< [x(t)]^n\right>$, $\dot{W}(t) =
\dot{k}(t) \left< [x(t)]^n\right>$, and, $\dot{Q}(t) = k(t) d\left<
    [x(t)]^n\right>/dt$, respectively. In order to describe the
    average thermodynamics, we
    thus have to consider the dynamics of the $n$th moment $\left<
    [x(t)]^n\right>$. Multiplying Eq.~\eqref{eq:Langevin_nonlinear} by
    $x^{n-1}$ and averaging the result over the noise, we find that
\begin{equation}
\frac{d}{dt}\left< [x(t)]^n\right> = - n k(t) \left< [x(t)]^{2n-2} \right>
+ n \left< [x(t)]^{n-1} \eta(t)\right>.
\label{eq:dx4}
\end{equation}
Thus, in order to get an exact closed dynamical equation for $\left< [x(t)]^n\right>$, we also need a dynamical equation for $\left< [x(t)]^{2n-2}\right>$ which, in turn, depends on the moment $\left< [x(t)]^{3n-4}\right>$, and so on. However, out of equilibrium each degree of freedom (and, also each moment $\left< [x(t)]^n\right>$) has, in general, its own effective temperature, if such a set of effective temperatures can consistently be defined at all. 

Recall that the development of a useful (finite-time) thermodynamic description
based on a time-dependent effective temperature requires a system with
equilibrium noise that yields the same (time-resolved) dynamics of the
relevant moments, for which our above discussion of the variance
of the linear model \eqref{eq:linear_model} provides the
paradigm. This means that we would have to develop a passive model
with an equilibrium noise  that gives rise to precisely the same
dynamics of all the moments in Eq.~(\ref{eq:dx4}) as the original nonlinear active model. \textcolor{black}{Even though our general discussion in Sec.~\ref{sec:GENEffT} shows that for the considered one-parameter potential (Hamiltonian) this should always be possible, this can get very difficult to achieve analytically~\cite{Caprini2019b} if the  moments
represent independent effective
degrees of freedom so that their effective temperatures differ}~\footnote{A way to overcome this limitation, leading to an approximate analytical effective temperature, might be based on finding a suitable (approximate) closure for Eq.~\eqref{eq:dx4}, so that it would only depend on a finite number of moments.}. \textcolor{black}{Nevertheless, in this case it should be possible to find the effective temperature numerically. As discussed in Sec.~\ref{sec:GENEffT}, for Hamiltonians that are not proportional to a single control parameter we are not able to give any general conclusions.}

Despite these limitations, there are also many important properties that can successfully be captured by the effective-temperature mapping. 
In the next section, we review its consequences for the performance of active heat engines.
Experts in stochastic thermodynamics may wish to continue directly with Sec.~\ref{sec:example}, where we derive and discuss more specific analytical results based on the so-called active Brownian particle (ABP) model with an exponential correlation matrix.  

\textcolor{black}{
\section{Linear theory: thermodynamics}
\label{sec:cotton}
}

\subsection{Effective entropy production}
\label{sec:equilirbium_model}

As described above, the dynamics of the variance in the active model
\eqref{eq:linear_model} can be mimicked exactly by the effective passive model in \eqref{eq:eq_model} with an equilibrium bath at the time-dependent temperature $T_{\rm
  eff}(t)$, as long as the latter does not transiently turn negative.
The noise intensity $D_{\rm eff}(t)$ and the mobility
$1/\mu$ in Eq.~\eqref{eq:eq_model} are thus related by the
fluctuation-dissipation relation $D_{\rm eff} = \mu T_{\rm eff}$.
Recall that the variance determines the
average thermodynamics of the active engine in terms of work, heat,
and efficiency. In particular, due to our interpretation of the thermodynamic variables,
the (average) performance of the active heat engine is the same as that of a passive heat engine based on
Eq.~\eqref{eq:eq_model} and can thus be taken over from the
known thermodynamics of classical heat engines~\cite{sch08,hol14,sek10}.
\textcolor{black}{In fact, such a (partial) thermodynamic framework based on the first and second law of thermodynamics 
is a crucial requirement for a consistent extension of the conventional notion of efficiency to conditions far from equilibrium.}

For pedagogical reasons and for completeness, we gather the explicit expressions that summarize the thermodynamics 
of the linear active heat engine, here. 
\textcolor{black}{The work reads
\begin{equation}\label{eq:linear_work}
W(t_{\rm i},t_{\rm f})  =\frac{1}{2}\int_{t_{\rm i}}^{t_{\rm f}}dt\, \underbrace{\dot{k}(t) \sigma(t)}_{\dot W(t)} = \frac{1}{2}\int_{k(t_{\rm i})}^{k(t_{\rm f})}dk\, \sigma,
\end{equation}
and the exchanged total heat is give by
\begin{equation}\label{eq:linear_heat}
Q(t_{\rm i},t_{\rm f}) =\frac{1}{2}\int_{t_{\rm i}}^{t_{\rm f}}dt\, \underbrace{k(t) \dot{\sigma}(t)}_{\dot Q(t)} = \frac{1}{2}\int_{\sigma(t_{\rm i})}^{\sigma(t_{\rm f})}d\sigma\, k.
\end{equation}
The cycle output work and input heat are still given by Eqs.~(\ref{eq:WH}) and (\ref{eq:qin}). Since $k>0$, the latter now explicitly reads 
\begin{equation}\label{eq:qin2}
  Q_\text{in}(t_{\rm p},0) =\frac{1}{2}\int_0^{t_{\rm p}}dt\, k \dot{\sigma}\Theta(\dot{\sigma}) 
\end{equation}
}
A main result (to
be derived below) is the explicit formulation of the second law of
thermodynamics in terms of the mapping to the passive model. 
It states that the active engine has a non-negative total effective (in the sense of the mapping to the passive model) entropy-production rate 
\begin{equation}
\dot{S}_{\text{tot}}^{\text{eff}}(t) = \mu T_{\text{eff}}(t)\,\sigma(t) \left[\frac{2}{\sigma(t)}- \frac{k(t)}{T_{\text{eff}}(t)}\right]^2 \ge 0.
\label{eq:total_RS_eff}
\end{equation}
Thermodynamically, the entropy production can always be decomposed into the contributions 
\begin{equation}
\label{eq:effRents}
\dot{S}_{\text{tot}}^{\text{eff}}(t) = \dot{S}^{\text{eff}}(t) + \dot{S}_{\text{R}}^{\text{eff}}(t) 
\end{equation}
due to the working substance itself and  due to the entropy
change in the (effective) heat bath, respectively.
Since, by definition, the heat flow from/into an equilibrium heat bath is
reversible, the entropy change of the bath obeys the Clausius equality,
\begin{equation}
\dot{S}_{\text{R}}^{\text{eff}}(t) = -\frac{\dot{Q}(t)}{T_{\rm eff}(t)}
= -\frac{k(t)\dot{\sigma}(t)}{2 T_{\rm eff}(t)}.
\label{eq:SRGauss}
\end{equation}
For the system entropy, one merely has the weaker Clausius inequality
\begin{equation}
\dot S^{\rm eff}(t) \geq - \dot S_{\rm R}^{\rm eff}(t) = \dot Q(t) /T_{\rm eff} (t)\,.
\label{eq:ClausiusIEQ}
\end{equation}
It can be turned into an equality if a quasi-static driving protocol
is employed, which then also optimizes the thermodynamic efficiency
of the active heat engine.

We now show how these results follow from the statistical-mechanics
description. First and foremost, note that the linearity of
Eq.~\eqref{eq:eq_model} ensures that the stochastic process $\textbf{r}(t)$ is a linear functional of the Gaussian white noise $\bm{\xi}(t)$. The probability density for the particle position $\textbf{r}=(x,y)$ at time $t$ is therefore also Gaussian, namely
\begin{equation}
p^{\rm eff}(x,y,t) = \frac{1}{\pi
  \sigma(t)}\exp\left[-\frac{(x^2+y^2)}{\sigma(t)}\right] \;,
\label{eq:PDFGauss}
\end{equation}
and can easily be seen to solve the Fokker-Planck equation
\begin{align}
\frac{\partial p^{\rm eff}}{\partial t}=\nabla_{\textbf{r}}\cdot\left[\mu \nabla_\textbf{r} \mathcal{V}(\textbf{r})+D_{\rm eff}\nabla_{\textbf{r}}\right]p^{\rm eff}
\label{eq:FPEeff}
\end{align}
with $\nabla_\textbf{r} = (\partial_x,\partial_y)$. 
The corresponding Gibbs-Shannon entropy
\begin{multline}
\label{eq:effentSys}
S^{\text{eff}}(t)= -  \int_{-\infty}^{\infty} dx  \int_{-\infty}^{\infty}dy\, p^{\rm eff} \log p^{\rm eff}\\=\log \sigma(t) + \log \pi + 1
\end{multline}
is thus solely determined by the variance $\sigma(t)$ of the
PDF~\eqref{eq:PDFGauss}, and therefore changes with the rate
\begin{equation}
\dot{S}^{\text{eff}}(t) = \frac{\dot{\sigma}(t)}{\sigma(t)}\,.
\label{eq:rateSGauss}
\end{equation}
The second law in the form given in Eq.~\eqref{eq:total_RS_eff} now
follows by inserting Eqs.~\eqref{eq:SRGauss} and \eqref{eq:rateSGauss}
into Eq.~\eqref{eq:effRents} and using Eq.~\eqref{eq:var_formal} for
the time derivative of the variance in the form
$\dot{\sigma} = 4\mu T_{\rm eff}\sigma (1/\sigma -k/2T_{\rm eff})$,
after rearranging the resulting terms.

To make the entropy production vanish, which corresponds to the equal sign in
Eqs.~(\ref{eq:total_RS_eff}) and (\ref{eq:ClausiusIEQ}),
one has to drive the engine quasi-statically.  This
amounts to setting $\dot{\sigma}=0$ in Eq.~\eqref{eq:Teffgen2}, which yields
\begin{equation}
\sigma(t) = \sigma_{\rm \infty}(t) \equiv 2T_{\rm eff}(t)/k(t) \,.
\label{eq:sigma_inf_EQ}
\end{equation}
For a quasi-static driving, the rates of change \eqref{eq:SRGauss} and \eqref{eq:rateSGauss} of the reservoir and system entropies also both vanish, since they are proportional to the vanishing time derivative $\dot{\sigma}=0$. However, this feature alone might not be enough for concluding that the entire entropy  
\begin{equation}
\Delta S^{\rm eff}_{\rm tot}(t_{\rm p}) = \int_0^{t_{\rm p}}dt' \dot{S}^{\rm eff}_{\rm tot}(t')
\label{eq:Stoteff}
\end{equation}
throughout the whole cycle vanishes as $t_{\rm p} \to \infty$, since it depends on how large $t_{\rm p}$ must be to ensure quasi-static conditions, which in turn depends on the intrinsic relaxation behavior of the working substance (in our case
the trapped colloid) \cite{lee2017}. It is a consequence of the fluctuation-dissipation relation fulfilled by the effective equilibrium model that the rates of change \eqref{eq:SRGauss} and \eqref{eq:rateSGauss} of the reservoir and system entropies converge to each other fast enough that the whole quasi-static cycle is reversible and \eqref{eq:Stoteff} vanishes for large $t_{\rm p}$. We come back to this issue in Sec.~\ref{sec:example}, where we analyze an explicit model realization.

\subsection{Efficiency bounds}
\label{eq:perfomance_limits}

For an arbitrary cycle, the Clausius inequality~\eqref{eq:ClausiusIEQ} can, 
via standard manipulations~\cite{Callen2006}, be rewritten in terms of the quasi-static (qs) bounds
for the output work  $W_{\rm out}$ and efficiency $\eta$, respectively,
\begin{gather}
W_{\rm out} \le  W^{\rm qs}_{\rm out},
\label{eq:WORKEQC}
\\
\eta  \le \eta^{\rm qs} \le \eta_{\rm C} = 1 - \frac{\min(T_{\text{eff}})}{\max(T_{\text{eff}})}.
\label{eq:effEQC}
\end{gather}
According to the discussion in the previous section, these conditions identically constrain the active heat engine.  Given any driving protocol for the variation of the control parameters $k(t)$ and $T_{\rm eff}(t)$, etc., along the cycle, the largest output work per cycle and the largest efficiency are thus attained for  quasi-static driving with $t_{\rm p} \to \infty$. The ultimate (Carnot) efficiency limit $\eta_{\rm C}$ for the active engine is thus reached in a quasi-static Carnot cycle composed of two ``isothermal'' branches, with constant $T_{\rm eff}$, interconnected by two ``adiabatic branches'', with constant entropy \eqref{eq:effentSys} and variance  $\sigma_{\infty}$.

Similarly, the mapping to the passive model~\eqref{eq:eq_model} implies that the finite-time performance of the active heat engine is the same as that of its effective passive replacement. For convenience, we summarize some consequences of this observation, here. 
The quasi-static conditions, needed to reach the upper bound $\eta_C$ on efficiency exactly, imply infinitely slow driving and thus vanishing output power. Naturally, such powerless heat engines are uninteresting for practical purposes \cite{Holubec2017}, where only finite-time processes are relevant, and, thus, other measures of engine performance have been proposed. A prominent role among them plays the maximum power condition. Schmiedl and Seifert \cite{sch08} showed that overdamped Brownian heat engines deliver maximum power if they operate in the so called low-dissipation regime \cite{Esposito2010a}. Their analysis implies that the efficiency at maximum power of the active heat engine is given by 
\begin{align}
\eta_{\text{MP}} = 1 - \sqrt{\frac{\min(T_{\text{eff}})}{\max(T_{\text{eff}})}}.
\label{eq:EMP}
\end{align}
This result applies if the engine is driven along a finite-time Carnot cycle composed of two isotherms of constant $T_{\rm eff}$ and two infinitely fast adiabatic state changes at constant $\sigma$, with a suitable protocol for the trap stiffness $k(t)$ that minimizes the work dissipated during the isothermal branches. We also note that the maximum-power condition was investigated for a specific class of active colloidal heat engines in Ref.~\cite{mar18}.

Actual technical realizations of heat engines are usually designed for a certain desired power output. Thus, even more useful than the knowledge of the efficiency at maximum power is the knowledge of maximum efficiency at a given power. Like the former, the latter is, for a Brownian heat engine of fixed design, attained when operating in the low-dissipation regime along a finite-time Carnot cycle \cite{Holubec2015,Holubec2016,Ma2018}. The exact numerical and approximate analytical value of the maximum efficiency at arbitrary power for our setting can be obtained using the approach of Ref.~\cite{Holubec2016}. Another universal result, applicable to the active Brownian heat engine, is
that, for powers $P$ close to the maximum power $P^\star$, the efficiency increases infinitely fast with decreasing $P$ (i.e. $|d\eta/d P|_{P \to P^\star}|  \to \infty$)~\cite{rya16,Holubec2016}. Therefore, it is usually advantageous to operate heat engines close to maximum power conditions [small $\delta P = (P^\star-P)/P^\star$], rather then exactly at these conditions ($\delta P = 0$)~\cite{Holubec2015}. Moreover, the results of Refs.~\cite{Holubec2015,Holubec2016,Ma2018} show that $\eta_{\rm C}$ can be attained only in the limit $\delta P\to 1$, where either the power $P$ completely vanishes, or it is negligible with respect to the maximum power $P^\star$. Recently, this insight led to a proposition of protocols yielding very large maximum power, thus allowing Brownian heat engines to operate close to (and practically with) Carnot's efficiency at large output power \cite{Holubec2017,Holubec2018}. As discussed in the following paragraph, active Brownian heat engines offer an alternative route for achieving this.

\textcolor{black}{
In the following sections and in App.~\ref{appx:PDF}, we explicitly analyze a specific realization of
an active heat engine to illustrate the merits and limitations of the
mapping to the ``passive dynamics''~\eqref{eq:eq_model}, with an equilibrium bath.}

\section{Worked example: the ABE model}
\label{sec:example}
\subsection{Model definition}

To exemplify the above findings for a specific model, we now consider the so-called ABP model. It is the standard minimal model for a particle embedded into an equilibrium bath at temperature $T$ but actively propelling with velocity $\textbf{v}(t)= v(t)\textbf{n}(\theta)$ in the direction determined by the diffusing unit vector $\textbf{n}(\theta)$ at angle $\theta(t)$.  Encouraged by experimental evidence \cite{wu00, kri16, zha17} and in accord with theoretical studies based on a rigorous elimination of (fast) active degrees of freedom \cite{mae14b, mae15}, the ABP model with harmonic confinement (Fig.~\ref{fig:HE}a) has recently also been used to model passive Brownian colloids embedded in an active bath (Fig.~\ref{fig:HE}b) \cite{ang11, har14, far15, zak17,Kumari2019}. Indeed, within the formalism \textcolor{black}{for a general additive noise} outlined above, the ABP model provides us with a simple realization of Eq.~\eqref{eq:linear_model} in terms of a trapped colloid driven by the non-equilibrium noise
\begin{equation}
\bm{\eta}= \sqrt{2 D(t)}\bm{\xi} + \textbf{v}(t)\,.
\label{eq:ABMnoise}
\end{equation}
Here the components of $\bm{\xi}=\left(\xi_x,\xi_y\right)$ are mutually independent zero-mean unit-variance Gaussian white noises, but the velocity term $\textbf{v}$
prohibits a straightforward equilibrium interpretation. It contributes an exponential term to the total noise correlation matrix
\begin{multline}\label{eq:2tcorr}
C_{ij}(t,t') = \left< \eta_i(t)\eta_j(t')\right> = \delta_{ij}\Bigg[2\sqrt{D(t)D(t')}\delta(t-t') \\ + \left. \frac{1}{2} v(t)v(t') \exp\left\{ -\int_{\text{min}(t,t')}^{\text{max}(t,t')}dt''\,D_{\text{r}}\left(t''\right)\right\}\right]\,.
\end{multline}
Such exponential memory has indeed also been found in a weak-coupling model for a passive tracer in an active bath \cite{mae14b,zwa73}. Besides, it is often employed as a tractable model for the complex correlations arising in strongly interacting systems. 

For the following, we assume that the translational diffusion coefficient $D(t)$ obeys the Einstein relation $D(t) = \mu   T(t)$, but do not constrain the rotational diffusion coefficient $D_{\text{r}}(t)$ in the same way. The latter describes the free diffusion of the particle orientation $\textbf{n}$ on a unit circle and is incorporated into the ABP equations of motion
\cite{fil12,ste16,mar16} through yet another independent zero-mean unit-variance Gaussian white noise  $\xi_{\theta}$, $\left<\xi_{\theta}(t)\xi_{\theta}(t') \right> = \delta(t-t')$. \textcolor{black}{The ABP equations then read:} 
\begin{eqnarray}
\dot{\textbf{r}}(t) &=& - \mu k \textbf{r}(t) + \textbf{v}(t)+ \sqrt{2 D(t)}\bm{\xi}(t)
\label{eq:Langevin1A}\\
\dot \theta(t) &=& \sqrt{2 D_{\text{r}}(t)}\xi_{\theta}(t)\,.
\label{eq:Langevin1B}
\end{eqnarray}
That the ABP model provides a proper non-equilibrium active noise, as desired for Eq.~\eqref{eq:linear_model}, is not only apparent from the two-time correlation matrix~\eqref{eq:2tcorr}, which fixes the average thermodynamics of the model in a way that is not consistent with a fluctuation-dissipation relation.  
It is further manifest in higher order correlation functions \cite{zhe13} that are sensitive to the non-Gaussian character of the noise~\eqref{eq:ABMnoise}. As illustrated in App.~\ref{appx:PDF}, this for example allows for a bimodal distribution of the coordinates $x$ and $y$,   so that the ABP model captures some of the generically non-Gaussian character of non-equilibrium fluctuations, lost in another widely employed active-particle model that represents the active velocity as an Ornstein-Uhlenbeck process \cite{sza14}. We note that these properties are essentially caused by the variable rotational noise $\xi_{\theta}$ and persist in a constant-speed ($v=\text{const.}\neq0$) version of the model. 

To emphasize the paradigmatic character of the heat engine described by the ABP-Eqs.~\eqref{eq:Langevin1A}--\eqref{eq:Langevin1B} with perdiodically driven parameters $k(t)$, $T(t)$, $v(t)$, $D_{\rm r}(t)$, we refer to it as the ABE model.  It involves three ingredients that can potentially drive it far from equilibrium: (i) If the stiffness $k(t)$ changes on time-scales shorter than the intrinsic relaxation time, the particle dynamics is not fast enough to follow the protocol adiabatically. (ii) If the rotational diffusion coefficient $D_{\rm r}$ is not constrained by the Einstein relation, the rotational degree of freedom can be considered connected to a second bath at a temperature distinct from $T$. In general, connecting a system to several reservoirs at different temperatures drives it out of equilibrium. (iii) Finally, the velocity term in the Langevin system is formally identical to a non-conservative force giving rise to persistent currents that prevent equilibration.

\subsection{Cyclic driving protocol}
\label{sec:driving}

\begin{figure}[t]
\begin{center}
\centering
\includegraphics[width=1.0\columnwidth]{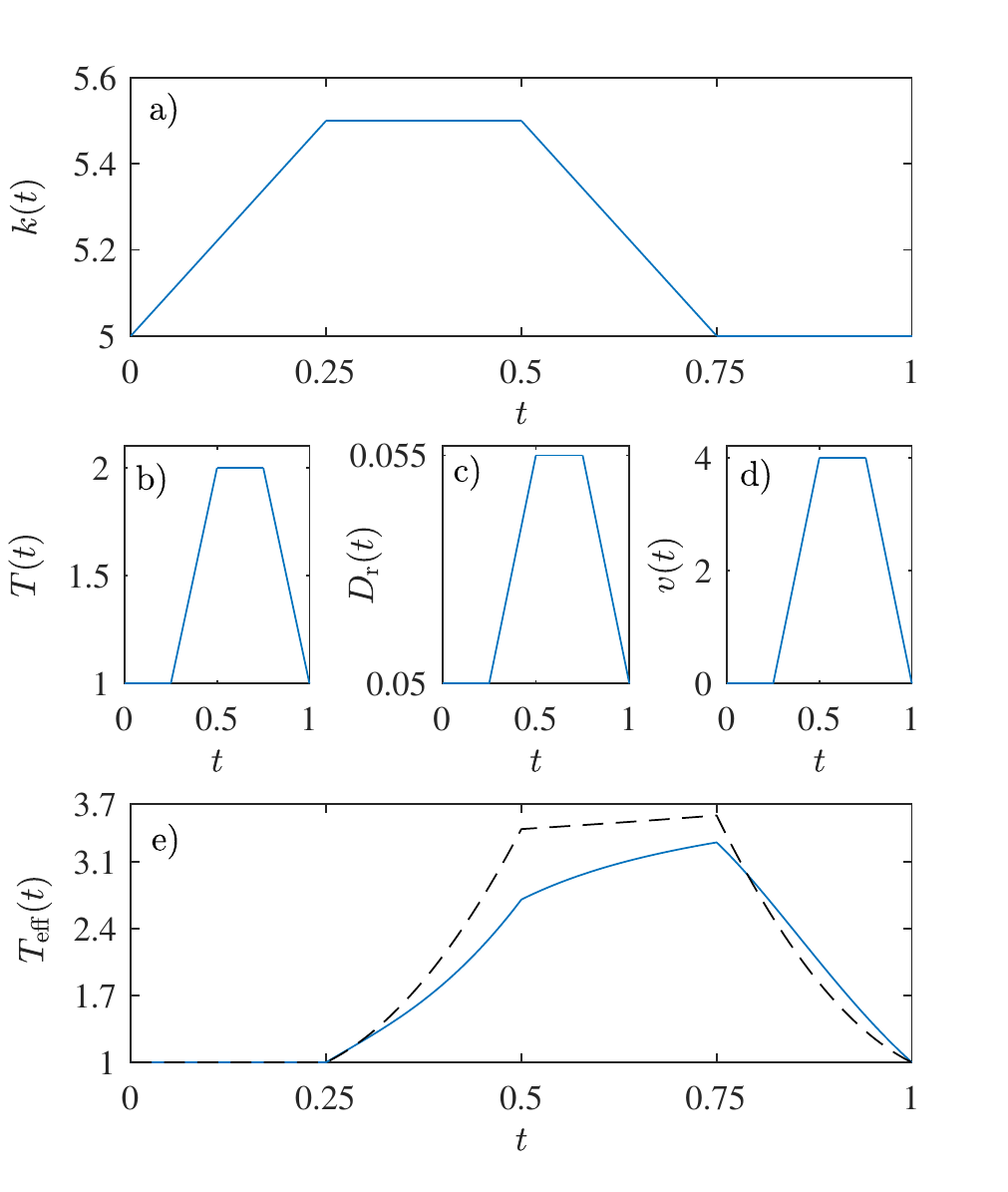}
\caption{The driving protocol of the ABE (a-d) and the effective temperature (e) that maps it to a passive model: a) trap stiffness, b) bath temperature, c) rotational diffusion coefficient, and d) active velocity, all as functions of time during the limit cycle. The full blue line in panel e) depicts the effective temperature $T_\textrm{eff}(t)$ of Eq.~(\ref{eq:effTsigma}), the dashed line its limit (\ref{eq:TeffInf}) for a quasi-static (infinitely slow) driving. Parameters used: $t_{\rm p}=1$, $k_< = 5$, $k_> = 5.5$, $T_> = 2$, $T_< = 1$, $D_{\text{r}}^> = 0.055$, $D_{\text{r}}^< = 0.05$, $v_>=4$, $v_< = 0$ and $\mu = 1$.}
\label{fig:driving}
\end{center}
\end{figure}

Our driving protocol involves a periodically modulated stiffness $k$, reservoir temperature $T$, rotational diffusion coefficient $D_{\text{r}}$, and active velocity $v$. We let the system evolve towards the limit cycle, where we analyze its performance. While the following theoretical discussion applies to arbitrary periodic driving, we exemplify our results with a specific Stirling-type protocol that mimics the experimental setup of Ref.~\cite{kri16} (see Fig.~\ref{fig:driving}). It consists of four steps of equal duration ($t_{\rm p}/4$):

(i) ``Isothermal'' compression $\rm{A}\to \rm{B}$:  the stiffness $k$ increases linearly from $k_<$ to $k_>$ at constant noise strength corresponding to the temperature $T = T_<$ and activity $\{D_{\text{r}},v\}=\{D_{\text{r}}^<,v_<\}$.

(ii) ``Isochoric'' heating $\rm{B}\to \rm{C}$: the noise strength $\{T,D_{\text{r}},v\}$ increases linearly from $\{T_<,D_{\text{r}}^<,v_<\}$ to $\{T_>,D_{\text{r}}^>,v_>\}$ at constant stiffness $k=k_>$. 

(iii) ``Isothermal'' expansion $\rm{C}\to \rm{D}$: the stiffness decreases linearly from $k_>$ to $k_<$ at constant noise strength $\{T_>,D_{\text{r}}^>,v_>\}$.

(iv) ``Isochoric'' cooling $\rm{D}\to \rm{A}$: the noise strength decreases back to its initial value at constant stiffness $k=k_<$.  

Note that the ``isothermal'' state changes are characterized by constant bath temperature and activity, which in general corresponds to a varying effective temperature [see Fig.~\ref{fig:driving}e)]. As explained in Secs.~\ref{sec:energetics} and \ref{sec:equilirbium_model}, the engine consumes (performs) work when $\dot{k} > 0$ ($\dot{k} < 0$), i.e. from A $\to$ B (C $\to$ D) as a standard Stirling engine. On the other hand, heat is absorbed (emitted) from (to) the reservoir when $\dot{\sigma} > 0$ ($\dot{\sigma} < 0$) and the corresponding portions of the cycle might be different than for the standard Stirling engine, depending on the behavior of the variance $\sigma$.

\subsection{Variance dynamics in the limit cycle}
\label{sec:variance}

\begin{figure}[t]
\begin{center}
\centering
\includegraphics[width=1.0\columnwidth]{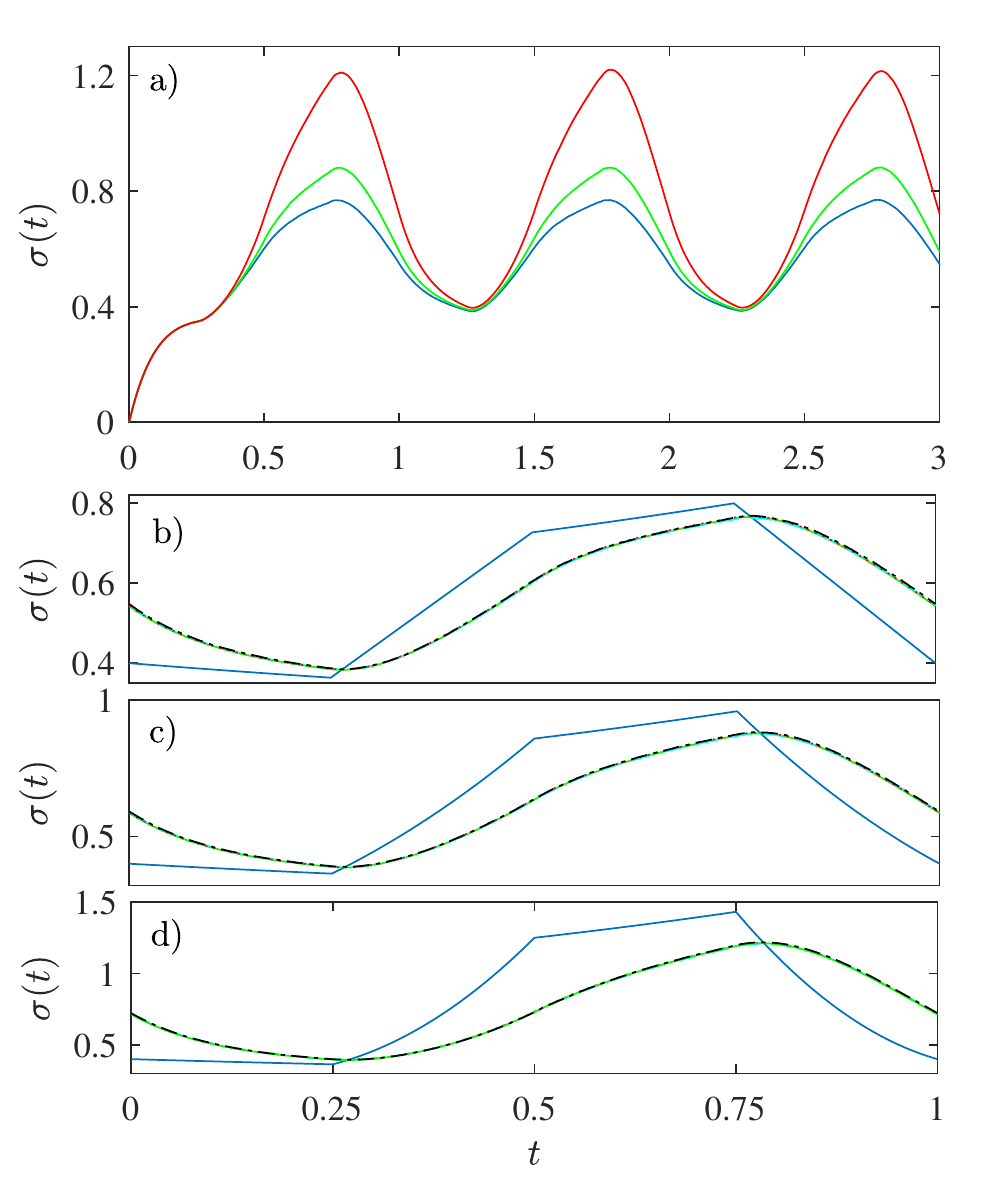}
\caption{Positional variance $\sigma(t) = \left<\mathbf{r}\cdot\mathbf{r} \right>$ over time for the protocol shown in Fig.~\ref{fig:driving}. Increasing activity $v_> = 0,2,4$ yields an increasing variance $\sigma$. a) Brownian Dynamics (BD) simulations of the relaxation to the limit cycle. b) c) d) The dynamics on the limit cycle in BD simulations (solid green), numerical solutions~\cite{hol18} (dot-dashed black), and from the analytical formula (\ref{eq:sigma_best}) (dotted red), shows perfect agreement, despite considerable distance from the quasi-static limit~\eqref{eq:VarInf} (broken blue lines).}
\label{fig:var}
\end{center}
\end{figure}

\textcolor{black}{
During the limit cycle, which is attained at late times, the dynamics of the variance $\sigma(t) = 2\sigma_x(t) = 2\sigma_y(t)$ [due to the symmetry of Eq.~\eqref{eq:Langevin1A}] is for arbitrary time-periodic driving governed by the two coupled ordinary differential equations
\begin{eqnarray}
\dot{H}(t) &=& - [\mu k(t) + D_{\text{r}}(t)] H(t) + v(t),
\label{eq:dynH}
\\
\label{eq:dyn}
\dot{\sigma}(t) &=& -2\mu k(t)\sigma(t) + 4D(t) + 2v(t)H(t)\,.
\end{eqnarray}
Here, the term $2D(t) + v(t)H(t)$ determines the long-time time-periodic behavior of the average $\left<\mathbf{r}(t)\cdot\bm{\eta}(t)\right>$. See App.~\ref{appx:var_der} for details of derivation of Eqs.~\eqref{eq:dynH}
 and \eqref{eq:dyn}. Their general solution reads 
\begin{eqnarray}
H&=&H_0 {\rm e}^{-F\left(t,0\right)}+\int_{0}^{t}dt'\, v\left(t'\right) {\rm e}^{-F\left(t,t'\right)},
\label{eq:H_best}\\
\sigma &=& \sigma_0 {\rm e}^{-2K\left(t,0\right)} + 4\int_0^t dt'\,D_{\text{eff}}(t'){\rm e}^{-2 K(t,t')}
\label{eq:sigma_best}
\end{eqnarray}
with functions $K(t,t_0)=\mu \int_{t_0}^{t} dt' k(t')$, $F(t,t_0)=K(t,t_0)+\int_{t_0}^{t} dt' D_{\text{r}}(t')$, and $D_{\text{eff}}(t)= D(t) + v(t) H(t)/2$. The constants  
\begin{eqnarray}
H_0 &=& \frac{\int_{0}^{t_{\rm p}}dt'\,v\left(t'\right){\rm e}^{-F\left(t_{\rm p},t'\right)}}{1-{\rm e}^{-F(t_{\rm p},0)}},\\
\sigma_0 &=& 4\frac{\int_0^{t_{\rm p}} dt'\,D_{\text{eff}}(t'){\rm e}^{-2 K(t_{\rm p},t')}}{1-{\rm e}^{-2K(t_{\rm p},0)}}
\end{eqnarray} 
secure the time-periodicity of the solution and thus they are fixed by the conditions $H(t_{\rm p}) = H(0)$ and $\sigma(t_{\rm p}) = \sigma(0)$.}

Quasi-static conditions correspond to slow driving relative to the relaxation times $\tau_H = 1/(\mu k + D_{\rm r})$ and $\tau_\sigma = 1/(2\mu k)$ for $H$ and $\sigma$, respectively. That allows the dynamics of the functions $H$ and $\sigma$ to be regarded as relaxed, $\dot{H}=\dot{\sigma}=0$, 
from which one gets the quasi-static variance 
\begin{equation}
\label{eq:VarInf}
\sigma(t)\to \sigma_\infty(t) \equiv\frac{2 }{k}\left(T+\frac{v^2}{2\mu  }\frac{1}{k\mu+D_{\text{r}}}\right)\,.
\end{equation}
The leading correction in the driving speed is derived in App.~\ref{appx:var_approx}. Conversely, if the driving is fast relative to the relaxation times $\tau_H$ and $\tau_\sigma$, the colloid cannot respond to the changing parameters $k$, $T$, $v$ and $D_{\rm r}$, and its variance is given by Eq.~\eqref{eq:VarInf} with time-averaged parameter values. 

At intermediate rates, the complete expression~\eqref{eq:sigma_best} has to be used. To make sure that we calculate the nested integral correctly, we cross-check the obtained results with two  independent methods, Brownian Dynamics (BD) simulations and numerical solutions~\cite{hol18}. The finite-time variances follow the quasi-static ones like carrot-chasing donkeys, i.e., the variance decreases (increases) if it is larger (smaller) than the stationary value $\sigma_\infty$ corresponding to the given value of the control parameters, cf.~Figs.~\ref{fig:var} (b)-(d). The discrepancy between the quasi-static and the finite-time predictions increases for faster driving and moreover grows with the activity ratio $v_>/v_<$. As intuitively expected, and suggested by the role of $v$ in Eq.~\eqref{eq:dyn}, larger active velocities lead to larger variances.

\subsection{Effective temperature}
\label{eq:eff_temp_ABP}

Comparing Eqs.~\eqref{eq:var_formal}, \eqref{eq:Teffgen2}, and \eqref{eq:dyn} we find for the effective 
temperature of the ABE on the limit cycle
\begin{equation}
T_{\text{eff}}(t) = \frac{D_{\text{eff}}(t)}{ \mu} = T(t) + \frac{v(t)H(t)}{2 \mu}.
\label{eq:effTsigma}
\end{equation}
Its value is always larger than the bath temperature $T$. Apart from the latter, it also depends on the activity $v$, mobility $\mu$, trap stiffness $k$, and rotational diffusion coefficient $D_r$. All the parameters, except for $T$, enter $T_{\text{eff}}$ indirectly, and in a complex way, through the differential equation~\eqref{eq:dynH} for $H$. The effective temperature thereupon acquires the characteristic relaxation time, $\tau_H = (\mu k + D_{\rm r})^{-1}$. Its quasi-static limiting form \eqref{eq:qs_Teff} explicitly reads
\begin{equation}
T_{\rm eff}(t) \to T_{\text{eff}}^{\infty}(t) \equiv \frac{k \sigma_{\infty}}{2} = T+\frac{v^2}{2\mu  }\frac{1}{k\mu+D_{\text{r}}} .
\label{eq:TeffInf}
\end{equation}

The effective temperature possesses several counter-intuitive features. First, in case of periodically modulated activity or trap stiffness,  it varies in time, even if the bath temperature is held constant. Moreover, due to its dynamical nature and finite relaxation time, it generally does so even when the parameters $T$, $v$, $D_{\rm r}$ and $k$ are held constant. Hence, to realize a proper (effectively) isothermal process with constant $T_{\rm eff}$, one has to carefully tune the control parameters. This is most easily achieved under quasi-static conditions, as demonstrated in Fig.~\ref{fig:driving}e). There we plot the effective temperature \eqref{eq:effTsigma} (full blue line) and also its quasi-static limit \eqref{eq:TeffInf}, which would be obtained at very slow driving (black dotted line). For the chosen parameters, the quasi-static effective temperature \eqref{eq:TeffInf} runs \emph{approximately} along a Stirling cycle, in accord with the temperature $T(t)$ and activity $v(t)$ [Figs. ~\ref{fig:driving} (b) -- (d)]. Conversely, the finite-time effective temperature \eqref{eq:effTsigma} exhibits substantially different behavior.

Before going into more details, we now outline three thermodynamically consistent interpretations of the ABE model and derive the corresponding entropy productions. In the discussion of quasi-static and finite-time performance of the engine in Secs.~\ref{sec:InfSlow} and \ref{sec:FTenergetics}, respectively, we utilize these entropy productions as examples of variables that are not captured by the effective-temperature mapping~\eqref{eq:eq_model}. Another example is the full distribution of the particle position, which we discuss in App.~\ref{appx:PDF}. 

\section{ABE entropy production}
\label{sec:entropyproduction}

As a genuinely non-equilibrium system, any active heat engine always produces entropy, even if operated infinitely slowly. However, how much of that entropy we can (or care to) track depends on our experimental resolution (and interpretation of the engine). 

\subsection{User perspective}

\begin{figure}[t]
\centering
\includegraphics[width=1.0\columnwidth]{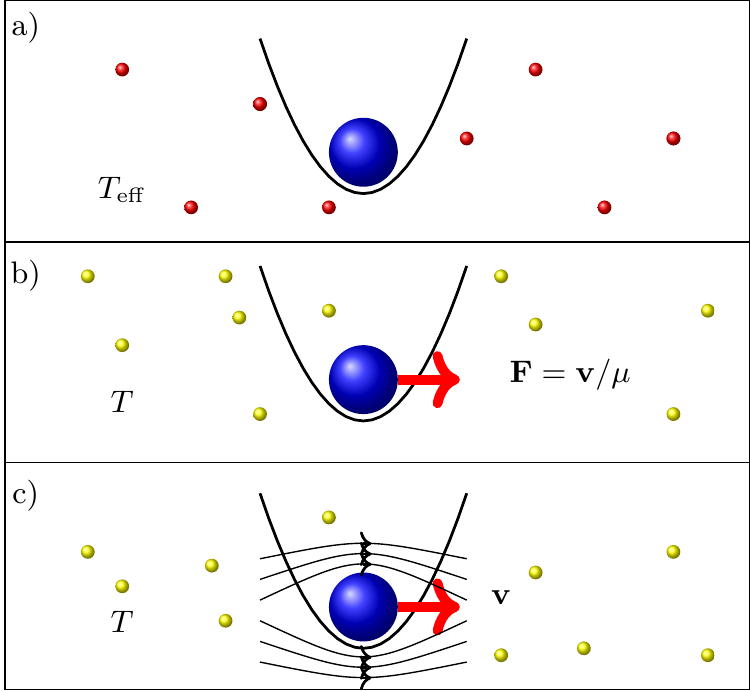}
\caption{Different levels of control over the system imply different changes in the bath entropy. a) Mere ``users'' of an active heat engine are only concerned with its thermodynamic input/output characteristics. They judge reversibility and entropy changes with respect to an effective equilibrium bath [red particles] at a fictitious temperature $T_{\rm eff}$ larger than the temperature $T$ of the background solvent [yellow particles in b) and c)]. More detailed knowledge about the engine's internal working substance (here the ABP particle) and its dynamics uncovers the non-equilibrium character of the system, which depends on the time-reversal properties of its dynamics~\cite{Dabelow2019,Crosato2019}.
If the active velocity $\textbf v$ in the ABE results from dragging or pushing the particle through the liquid by an external force $\textbf F$ [panel b)], the particle behaves like a sailboat and the change in bath entropy obeys Eq.~\eqref{eq:Sreven}. If the particle is self-propelled or advected by the surrounding liquid with velocity $\textbf v$ like a surfboard [panel c)], the bath entropy obeys Eq.~\eqref{eq:Srodd}.}
\label{fig:entropysketch}
\end{figure}

On the coarsest level of description, which might be adopted by a \emph{user} of the heat engine, only the supplied heat and the harvested output work matter.  Their ratio is the natural measure of efficiency, which is bounded by the optimum (Carnot) efficiency determined by the effective temperature $T_\text{eff}$. As we have discussed, this temperature can experimentally be measured for the model of a trapped Brownian particle, namely by a device sensible to the variance $\sigma$ of the particle position; see Fig.~\ref{fig:entropysketch}~a). The thermodynamics of the active heat engine is thereby mapped to that of an ordinary engine with an equilibrium bath and obeys the same limitations. Accordingly, the user would conclude that the total dissipated cycle entropy 
\begin{equation}
 \Delta S_\text{tot}^\text{eff} = \int_0^{t_\text{p}} dt \dot S_\text{tot}^\text{eff} =
\int_0^{t_\text{p}} dt \dot S_\text{R}^\text{eff}
\end{equation}
is given by the net entropy change per cycle in the bath, which thus solely controls the degree of irreversibility of the cycle. To compute the latter, the user would resort to the expression given in Eq.~\eqref{eq:SRGauss} of Sec.~\ref{sec:equilirbium_model}, namely
\begin{equation}
  \dot S_R^\text{eff} = 
  - \dot{Q}/T_\text{eff} \equiv  \dot Q^\text{eff}_\text{dis}/T_\text{eff} \,.
 \label{eq:SREff}
\end{equation}
Since the particle dynamics is modelled within an overdamped Stokes approximation, the corresponding ``effective'' dissipation $Q^\text{eff}_\text{dis}$ to the effective equilibrium bath is straightforwardly given by the force acting on the particle times its velocity (averaged)
\begin{equation}
\dot Q^\text{eff}_\text{dis} = -\langle \nabla_\textbf{r} {\mathcal H}\cdot \dot{\textbf{r}} \rangle = -k(t)\dot \sigma(t)/2.
\end{equation}

Importantly, the user is not concerned with other details of the non-equilibrium bath than the variance $\sigma$ and the effective equilibrium temperature $T_{\rm eff}$ it provides. He would thus adopt the above expressions for arbitrary noise in Eq.~\eqref{eq:linear_model}, regardless of the underlying physics of the bath. For the specific ABE realisation of the active heat engine, these expressions can explicitly be evaluated using Eqs.~\eqref{eq:dyn}, \eqref{eq:sigma_best} and \eqref{eq:effTsigma}. \textcolor{black}{This notion of entropy production, directly derived from the notion of system entropy consistent with the second law for the supplied heat and the harvested output work, is the only one to safely yield efficiency bounds compatible with conventional definitions. It is thus arguably the most pertinent one in the context of active heat engines.}

\subsection{Trajectory perspective}

In contrast to the above user, a \emph{heat engineer} would possibly consider the engine at a higher resolution and have access to the individual stochastic \emph{trajectories} of the particle position generated by Eq.~\eqref{eq:linear_model}.  Thereby, she could  uncover the non-equilibrium character of the active heat bath, which dissipates energy even if the engine operates under quasi-static conditions. To this end, she could evaluate the dissipation per cycle in the form $\left<\log P_{\text F}(\Gamma)/P_{\text R}(\Gamma^\star)\right>$, exploiting a relation often referred to as local detailed balance condition. It relates the symmetry breaking between the path probabilities $P_{\text F}(\Gamma)$ and $P_{\text R}(\Gamma^\star)$ for paths $\Gamma$ and their time-reversed images $\Gamma^\star$ to dissipation.  (For more details, see App.~\ref{appx:entropy} and Refs.~\cite{mae03, Battle2016}.) The method can in principle be applied regardless of the physics underlying the noise term in Eq.~\eqref{eq:linear_model}, if one can observe or otherwise guess the time-reversed dynamics. See Ref.~\cite{Battle2016} for an example of a successful application of such a strategy to biological systems. In general, this will however technically require assumptions or knowledge of the time-reversed noise dynamics, i.e., microscopic information beyond that of the stochastic (forward) trajectories of the particle position. Such information is seldom available outside the realm of detailed models of the mesoscopic physics. For specificity, we therefore now consider explicitly the ABE model, based on the concrete ABP model.

\subsection{ABP perspective: sailboats versus surfboards}
\label{sec:surfsandsails}

For ABP's, the noise $\bm \eta$ comprises the (time-symmetric) equilibrium white noise $\sqrt{2D}\xi$ together with the active propulsion $\textbf{v}$. The colloid could be a randomly (self-)~propelled active particle or a schematically modeled passive tracer in an active bath \cite{kri16,zak17}. In any case, its active velocity $\textbf{v}$ is due to a dissipative process and admits two alternative interpretations, depending on its presumed time-reversal properties~\cite{Dabelow2019,Crosato2019}.  
Namely, it can be understood as a Stokes velocity caused by an external (random) force $\textbf{v}(t)/\mu$, the so-called swim force. 
This very common interpretation, depicted in Fig.~\ref{fig:entropysketch}) b), treats the particle like a sailboat blown around by erratic winds, which is why we refer to it as the ``sailboat'' interpretation. Or, in a second interpretation, depicted in Fig.~\ref{fig:entropysketch}) c), the active term $\textbf{v}(t)$ can be interpreted as the actual swim velocity of a microswimmer that either ``sneaks''  through the quiescent background solvent by an effective phoretic surface slip $\textbf{v}(t)$~\cite{Qian.2013,Kroy2016,khadka2018} or is passively advected by a local flow field $\textbf{v}(t)$~\cite{Speck2008,fod16}. We refer to it as the ``surfboard'' interpretation. It treats $\textbf{v}(t)$ as a proper dynamic velocity as opposed to the disguised force in the sailboat interpretation. Upon time-reversal, 
forces usually do not change the sign, while velocities do. The detailed balance condition then implies that the rate of entropy change in the bath reads 
\begin{equation}
\dot  S_{\text{R}}^{\pm} =   \dot Q_\text{dis}^\pm/T\,,
\end{equation}
for sailboats ($+$) and surfboats ($-$), respectively~\cite{cha16, pie17b, Marchetti2018}. The corresponding dissipation rates are
\begin{equation}
\dot Q^+_\text{dis} =  \langle (\textbf{v}/\mu - \nabla_\textbf{r} {\mathcal H}) \cdot \dot{\textbf{r}} \rangle = \dot Q^\text{eff}_\text{dis} +v^2\!/\mu -\langle \nabla_\textbf{r} {\mathcal H}\cdot {\textbf{v}}\rangle\,,
\label{eq:Qdis+}
\end{equation}
\begin{equation}
  \dot  Q^-_\text{dis} = \langle -  \nabla_\textbf{r} {\mathcal H}\cdot (\dot{\textbf{r}} -\textbf{v})\rangle = 
  \dot Q^\text{eff}_\text{dis} + \langle \nabla_\textbf{r} {\mathcal H}\cdot {\textbf{v}} \rangle\,.
\end{equation}
We refer to App.~\ref{appx:entropy} for details of the formal derivation, and discuss these results on a physical basis. In the dissipation rate $\dot Q^+_\text{dis}$ for sailboats, the swim term is added as an additional force (intuitively the wind drag) to the potential force. In contrast, for surfboards, it is subtracted from the particle velocity corresponding to a reformulation of the equation of motion in a frame that is freely co-moving with the flow velocity $\textbf{v}(t)$.


Since $\dot Q_\text{dis}^+(t)$ and $\dot Q_\text{dis}^-(t)$ have different reference points (vanishing for sailboats blown against the quay and  surfboards floating freely with the surf, respectively), the two dissipation rates can not generally be ordered according to their magnitude for the ABE, where both situations may (approximately) be encountered along the cycle. Also note that the detailed balance condition imposes that the heat is dissipated in the background solvent at temperature $T(t)$, which is natural from the point of view of the ABP model. As a consequence, also different amounts of entropy production will be assigned to the self-propulsion, dependent on the chosen ABP interpretation. 

They can both be understood as composed of the effective dissipation $Q_\text{dis}^\text{eff}(t)$ over the solvent temperature $T(t)\leq T_\text{eff}(t)$, plus some extra (manifestly active) entropy production due to the particle's excursions off the surf or off the quay, respectively,  
\begin{equation}
   T \dot S_{\text{R}}^{+} = 
  \dot{Q}_\text{dis}^\text{eff} +v^2/\mu - 2\mu k (T_{\rm eff}-T) \,,
\label{eq:Sreven}
\end{equation}
\begin{equation}
T\dot S_{\text{R}}^{-} =   \dot{Q}_\text{dis}^\text{eff} + 2 \mu k (T_{\rm eff} - T)\,,
\label{eq:Srodd}
\end{equation}
were we used $\langle \nabla_\textbf{r} {\mathcal H}\cdot {\textbf{v}} \rangle = k \langle \textbf{r} \cdot \textbf{v} \rangle = 
k \langle \textbf{r} \cdot (\bm{\eta} - \sqrt{2 D(t)}\bm{\xi}) \rangle = 2\mu k (T_{\rm eff}-T)$, which follows from Eqs.~\eqref{eq:Teffgen} and \eqref{eq:ABMnoise}. In the second case (surfboards), the additional propulsion contribution to the entropy production beyond $\dot S_{\text{R}}^\text{eff}$ is manifestly positive, since $T_\text{eff}\geq T$. Intuitively, this is because any failure to float with the flow gives rise to dissipation. In the first case (sailboats), the minimum condition for $\dot S_{\text{R}}^\text{eff}$ can only be guaranteed under quasi-static conditions. Intuitively, the ``wind'' may otherwise transiently prevent dissipation by ``arresting the sailboat at the quay''.

While the derivation of the expressions \eqref{eq:Sreven} and \eqref{eq:Srodd} relies on a deeper knowledge of the system dynamics than the behavior of the variance, it is worth noting that $\sigma(t)$ is still sufficient for their evaluation.  The dynamics of the variance thus suffices to evaluate the ``total'' entropy $\Delta S_{\rm tot}^\pm(t_{\text p}) = S_{\text R}^\pm(t_{\text p}) = \int_0^{t_{\text p}} dt\,\dot{S}_{\text R}^\pm$ produced per cycle of the operation of the ABE. In contrast, the change in the system entropy
\begin{equation}
\label{eq:sys_ent}
S(t) = -  \int_{-\infty}^{\infty} dx  \int_{-\infty}^{\infty}dy \int_0^{2\pi} d\theta\, p \log p\,,
\end{equation}
which vanishes for a complete cycle but is necessary for evaluating the total entropy change within the cycle, $\Delta S_{\rm tot}^\pm(t) = S_{\text R}^\pm(t) + S(t) - S(0)$, depends on the full probability distribution $p(x,y,\theta,t)$ for the position of the particle at time $t$. The latter obeys the Fokker-Planck equation
\begin{align}
\frac{\partial p}{\partial t}=\left(\nabla_\textbf{r} \cdot\left[\mu \nabla_\textbf{r} \mathcal{H}(\textbf{r}) - \textbf{v}\right]+D\nabla^2_\textbf{r}+D_{\rm r}\frac{\partial^2}{\partial\theta^2}\right)p
\label{eq:FPE}
\end{align}
corresponding to Eqs.~\eqref{eq:Langevin1A} and \eqref{eq:Langevin1B}. One can calculate the PDF $p(x,y,\theta,t)$ either numerically, from Eq.~\eqref{eq:FPE}, or using BD simulations of Eqs.~\eqref{eq:Langevin1A}--\eqref{eq:Langevin1B} (see App.~\ref{appx:PDF} for a detailed discussion of the results). The system entropy $S(t)$ is thus the only variable of our thermodynamic analysis which generally cannot be calculated using the mean square displacement $\sigma$ alone. 

The above results are suitable to fully quantify the engine's thermodynamic performance. In the following section we evaluate the derived expressions and discuss their generic properties.

\section{ABE-performance}\label{sec:ABE-performance}

\begin{figure}[t]
\centering
\includegraphics[width=1.0\columnwidth]{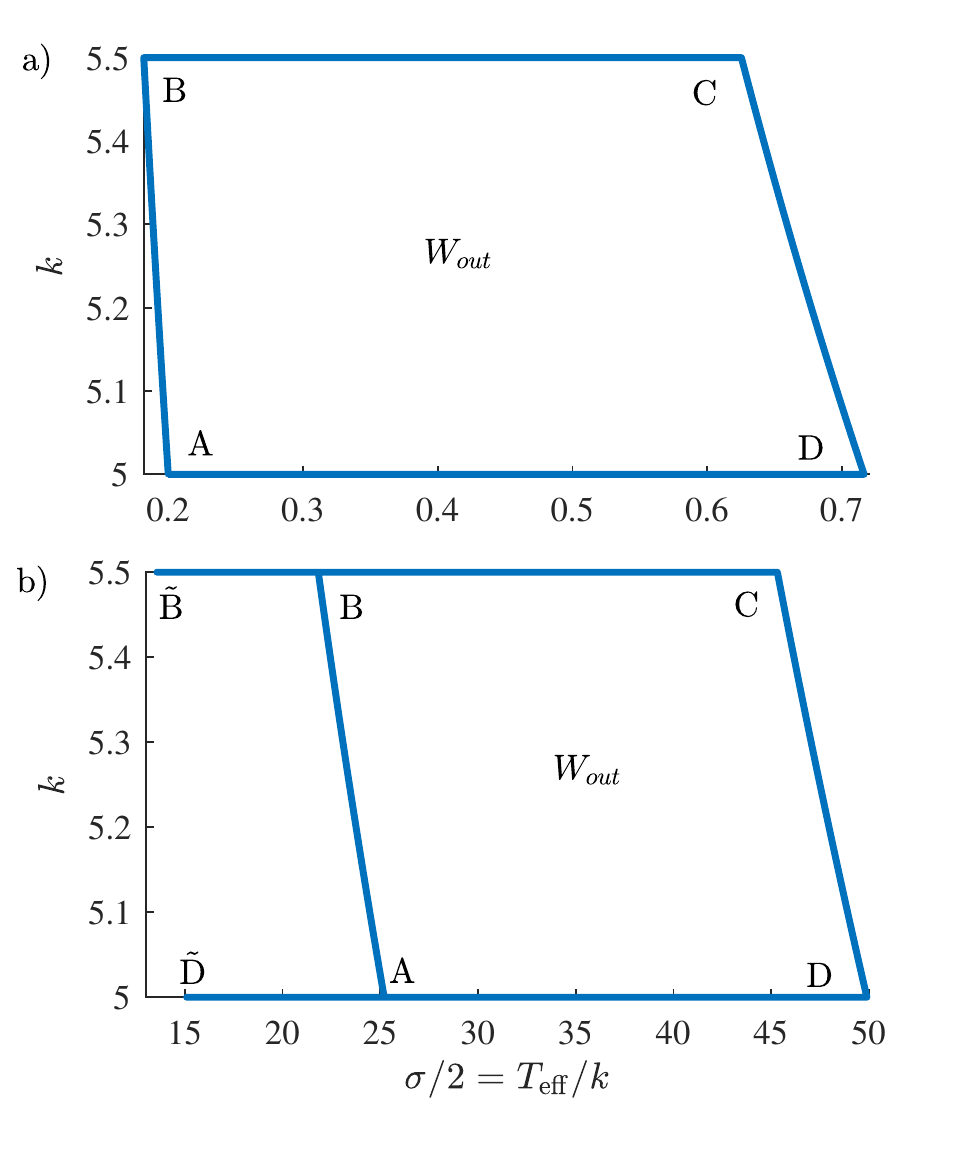}
\caption{Two quasi-static (generalized) Stirling cycles in terms of the trap stiffness $k(t)$ and variance $\sigma (t)$ of the particle positions ABCD and AB$\tilde{\rm B}$CD$\tilde{\rm D}$. (The corresponding energy flows are evaluated in Fig.~\ref{fig:energetics_QS}.) a) In the standard Stirling cycle ABCD, the heat flows from the bath into the system along the isochor BC and isotherm CD ($\dot{Q} = k \dot{\sigma}/2 > 0$), and from the system into the bath otherwise ($\dot{Q} < 0$). b) In the ``nonstandard Stirling'' cycle AB$\tilde{\rm B}$CD$\tilde{\rm D}$, the heat flow reverses (outflow along B$\tilde{\rm B}$, inflow along $\tilde{\rm B}$C) along the isochoric branch BC = B$\tilde{\rm B}$C and similarly for the isochor DA = D$\tilde{\rm D}$A. The output works $W_{\rm out} = \int_0^{t_{\rm p}}\sigma dk/2$ of the individual cycles are given by the areas they enclose. Similarly, heat input and output can be visualized as areas below the curves.}
\label{fig:diagram}
\end{figure}

In this section, we first focus on the quasi-static regime of operation of the ABE, where we demonstrate in more detail some peculiarities connected with the unintuitive behavior of the effective temperature. For vanishing entropy productions $\Delta S_{\rm tot}^\pm$, as defined in the previous section, the non-equilibrium ABE bath is seen to admit a representation as an equilibrium bath. Then, we consider finite-time effects onto the performance of the ABE, and the additional entropy production due to the non-quasi-static operation.


\subsection{Quasi-static regime}
\label{sec:InfSlow}

\begin{figure}[t]
\begin{center}
\centering
\includegraphics[width=1.0\columnwidth]{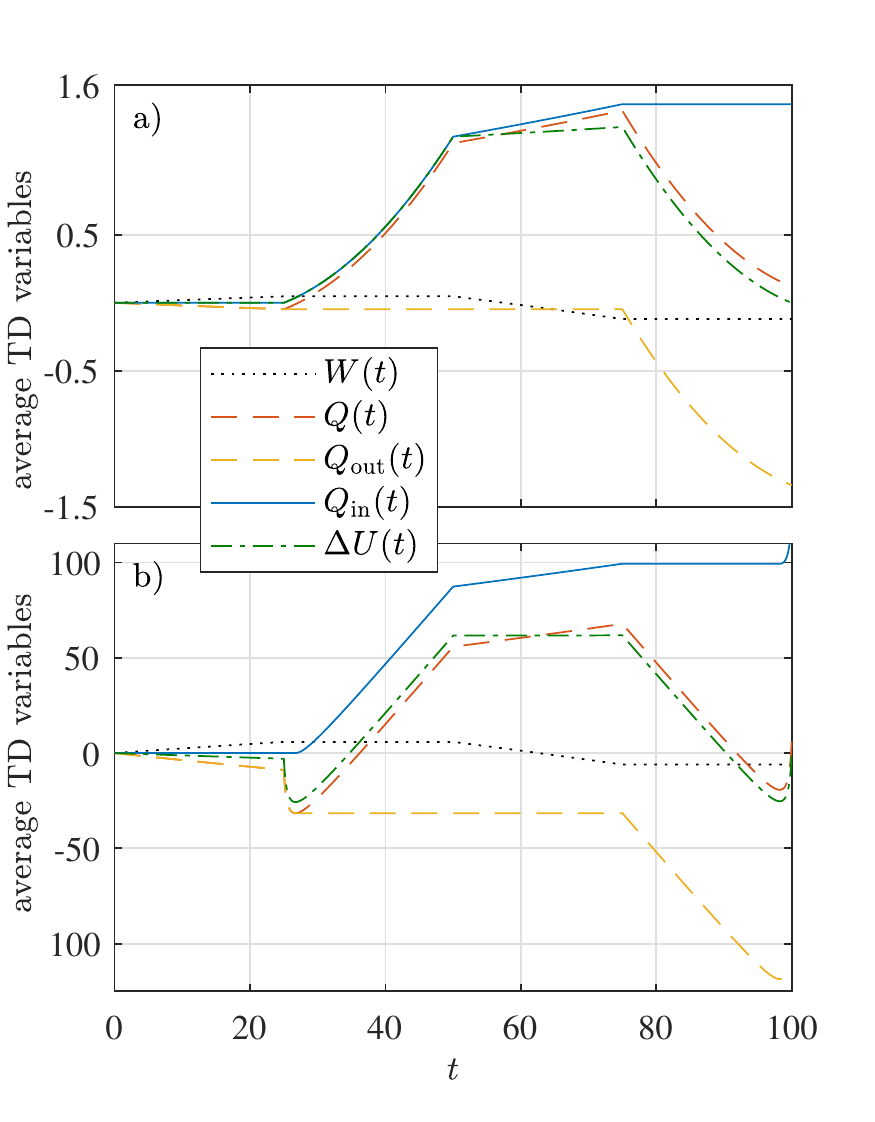}
\caption{Energy flows for the quasi-static Stirling cycles depicted in Fig.~\ref{fig:diagram}. Net work and heat $W$, $Q=Q_{\text{in}}+Q_{\text{out}}$, heat influx and outflux $Q_{\text{in}}$, $Q_{\text{out}}$, and internal energy change $\Delta U \equiv \langle \mathcal H(t) - \mathcal H(0)\rangle $, as defined in Secs.~\ref{sec:energetics} and \ref{sec:equilirbium_model}, are traced out as functions of time during a quasi-static cycle of duration $t_{\rm p} = 100$ significantly larger than the relaxation times $\tau_H = 1/(\mu k + D_{\rm r})$ and $\tau_\sigma = 1/(2\mu k)$ for $T_\text{eff}$ and $\sigma$, respectively. Panel a) $v_> = 4$. Panel b) $v_>= 500$, $v_<= 50$, $D_{\text{r}}^> = 500$ and $D_{\text{r}}^<= 5$; other parameters as in Fig.~\ref{fig:driving}.}
\label{fig:energetics_QS}
\end{center}
\end{figure}

In the quasi-static regime, the engine dynamics in terms of the variance $\sigma(t)$ and the effective temperature $T_{\rm eff}(t)$ are given by Eqs.~\eqref{eq:VarInf} and \eqref{eq:TeffInf}, respectively. They thus depend merely  parametrically on the driving $k(t)$, $T(t)$, $D_{\rm r}(t)$, and $v(t)$. The effective entropy production $\Delta{S}^{\rm eff}_{\rm tot}$~\eqref{eq:total_RS_eff} then vanishes, and the (effective) efficiency of the ABE is given by the classical result evaluated in terms of the stiffness $k(t)$ and temperature $T_{\rm eff}(t)$.   In particular, a quasi-static cycle consisting of two branches with constant $T_{\rm eff}$ and two adiabats will thus operate with Carnot efficiency $\eta_{\text C}$, \eqref{eq:effEQC}. Equivalently, realizing a Stirling cycle in terms of $k(t)$ and the effective temperature $T_{\rm eff}(t)$ will result in the  (effective)  Stirling efficiency $\eta_{\text C} \log a/(\eta_{\text C} + \log a)$ with $a = \text{min}(k) / \text{max}(k) $ \cite{zak17}. And one could deal similarly with other thermodynamic cyclic protocols. However, using the simplifying analogy with the effective equilibrium bath, one should make sure to actually use $k(t)$ and $T_{\rm eff}(t)$ as control parameters and not simply rely on an intuition about the behavior of the effective temperature based on the background solvent temperature $T$, activity $v$ and rotational diffusivity $D_{\rm r}$. Indeed, as mentioned in Sec.~\ref{eq:eff_temp_ABP}, what is a Stirling (or Carnot) cycle in terms of the effective temperature can be quite different from the one defined in terms of $T$, $v$, and $D_{\rm r}$. 
To quantify the difference, it is useful to introduce the parameter 
\begin{equation}
\mathcal{K}(t) \equiv k\mu /D_{\text{r}}
\label{eq:Kcal}
\end{equation}
which compares the characteristic timescales $D_{\rm r}^{-1}$ and $(k\mu)^{-1}$ for relaxation of the orientation $\theta$ and the position $\mathbf{r}$, respectively. The quasi-static effective temperature~\eqref{eq:TeffInf} can be written as
\begin{equation}
T_{\rm eff}(t) = T + \frac{v^2}{2\mu   D_{\rm r}}\frac{1}{1+\mathcal{K}}.
\end{equation}
Only in the limiting cases $\mathcal{K}\to 0$ and $\mathcal{K}\to \infty$, a \emph{naive} quasi-static isothermal process (constant temperature $T$, activity $v$, and rotational diffusivity $D_{\rm r}$ and variable stiffness $k$) corresponds to an effective equilibrium isothermal process (constant $T_{\rm eff}$). 

Despite the equilibrium analogy, the bath actually corresponds to a driven system with its apparent equilibrium characteristics actively maintained by some dissipative processes. So even for quasi-static operation of the engine, closer inspection reveals this non-equilibrium nature of the bath. \textcolor{black}{In particular, the sailboat/surfboard interpretations of the particle motion will reveal some of this entropy production, since 
\begin{equation}
   T \dot S_{\text{R}}^{\pm} = 
  \dot{Q}_\text{dis}^\text{eff} +\frac{v^2}{\mu} \frac{1}{1 + \mathcal{K}^{\pm 1}} \ge \dot{Q}_\text{dis}^\text{eff} 
\end{equation}
and thus
$\Delta{S}^{\pm}_{\rm tot} \ge \Delta{S}^{\rm eff}_{\rm tot}=0$.}

\textcolor{black}{
For strong confinements, $\mathcal{K}\gg1$, the active dynamics is highly persistent on the confinement scale, so that the particle moves quasi-ballistically in the potential. The effective temperature $T_{\rm eff}$ is therefore given by the temperature of the equilibrium solvent $T$, which is the only remaining source of noise. Using the sailboat interpretation of the ABP (for which $\mathbf{v}$ is interpreted as as an external force), we find that $\Delta{S}^+_{\rm tot} = \Delta{S}^{\rm eff}_{\rm tot} = 0$ since the sailboat is trapped in a quay. The sailboat interpretation is thus consistent with the notion that the ABE operates reversibly. In contrast, a trapped surfboard (for which $\mathbf{v}$ is interpreted as a velocity) is inhibited from moving with the surf, leading to 
dissipation:  $\Delta{S}^-_{\rm tot} = \Delta{S}^{\rm eff}_{\rm tot} + \int_0^{t_{\rm p}}dt\,v^2(t)/\mu = \int_0^{t_{\rm p}}dt\,v^2(t)/\mu>0$.}


\textcolor{black}{
For weak confinements, $\mathcal{K}\ll 1$, the particle's active motion randomizes on the confinement scale so that it can be subsumed into the $\delta$-correlated  noise \eqref{eq:ABMnoise} via the effective temperature and the corresponding noise correlation matrix $C_{ij}(t,t')= 2 \sqrt{D_{\rm eff}(t)D_{\rm eff}(t')} \delta_{ij}\delta(t-t')$. Its dynamics mimics Brownian motion in an effective equilibrium bath maintained at the (stiffness-independent) temperature $T_{\rm eff} =T + v^2/(2\mu D_r)$. In this case, confinement and random active motion interfere in such a way that both the sailboat and surfboard interpretations can detect the positive entropy production, $\Delta{S}^\pm_{\rm tot} > \Delta{S}^{\rm eff}_{\rm tot} = 0$, and the actual irreversibility of the operation. Only by imposing the additional limit $v^2 \ll 2\mu D_{\rm r} T$, when the rotational motion completely obliterates the active swimming so that $T_{\rm eff} = T$, surfboards cease to be bothered by the confinement and no longer dissipate, i.e., $\Delta{S}^-_{\rm tot} = 0$. In the sailboat interpretation, the release of the boat from the tug of war with the quay instead results in a complete waste of the efforts of the external swim force to haul the particle around in an enhanced random motion. The corresponding dissipation of the fully released sailboat thus precisely matches that of a fully trapped surfboard: $\Delta{S}^+_{\rm tot} = \int_0^{t_{\rm p}}dt\,v^2(t)/\mu>0$.}

For intermediate values of $\mathcal{K}$, the effective temperature depends on the stiffness $k(t)$ and the (traditional) definition of heat input along an individual step of the driving protocol may not actually yield the correct interpretation. It then also fails to yield a consistent measure of efficiency. Instead, one should carefully reconsider what is the actual heat input, based on Eq.~\eqref{eq:qin2}. Heat thus flows into the system whenever the variance $\sigma$ --- and thus the effective system entropy \eqref{eq:effentSys} --- increases, and vice versa.

To illustrate this point, recall the definition of the Stirling cycle in Sec.~\ref{sec:driving}. The standard Stirling cycle consists of two isochores (constant trap stiffness $k$) and two isotherms (constant solvent temperature $T$). Therefore it forms a rectangle in a $k$-$T$ diagram, translating to a shape similar to the ABCD cycle in Fig.~\ref{fig:diagram}, in a $k$-$T/k$ diagram. Actually, Fig.~\ref{fig:diagram} is slightly more general, as it shows two possible interpretations of the quasi-static ABE-Stirling cycle in a $k$-$T_\text{eff}/k$ diagram. The ``standard'' protocol ABCD corresponds to the evolution of the thermodynamic variables as depicted in Fig.~\ref{fig:energetics_QS}a). Note that they, in turn, evolve strictly monotonically or remain constant during the individual steps of duration $t_{\rm p}/4$. Hence, during a single step, heat is either only absorbed or only released by the system, and it is possible to write the input heat as $Q_{\rm in} = Q_{\text{BC}} + Q_{\text{CD}}$, where $Q_{\text{XY}}$ is the amount of heat absorbed between the points $X$ and $Y$. Which corresponds to the conventional practice for a Stirling cycle.

Consider next the cycle AB$\tilde{\rm B}$CD$\tilde{\rm D}$ corresponding to Fig.~\ref{fig:energetics_QS}b). In this case, the system releases heat during the segment ${\rm B}\tilde{\rm B}$  ($\sigma$ decreases from $\sigma_{\rm B}$ to $\sigma_{\rm \tilde{\rm B}}$), but absorbs heat during the remainder of the state change BC ($\sigma$ increases from $\sigma_{\tilde{\rm B}}$ to $\sigma_{\rm C}$).  A similar situation occurs also at the end of the cycle. Hence, the conventional shorthand notion of heat input as heat exchanged between the system and the reservoir during an entire step of the cycle is not appropriate, in this case. Instead, one has to use the definition ~\eqref{eq:qin}, also utilized in Fig.~\ref{fig:energetics_QS}. The dashed red, dashed yellow and full blue lines in Fig.~\ref{fig:energetics_QS}b) in the time interval from $t=25$ to $t = 50$, also serve to illustrate the differences in the heat balance. For a further treatment of efficiency of Stirling engines operating in contact with active baths in the quasi-static regime, we refer to Ref.~\cite{zak17}.

\subsection{Finite-time performance}
\label{sec:FTenergetics}

\begin{figure}[t]
\begin{center}
\centering
\includegraphics[width=1.0\columnwidth]{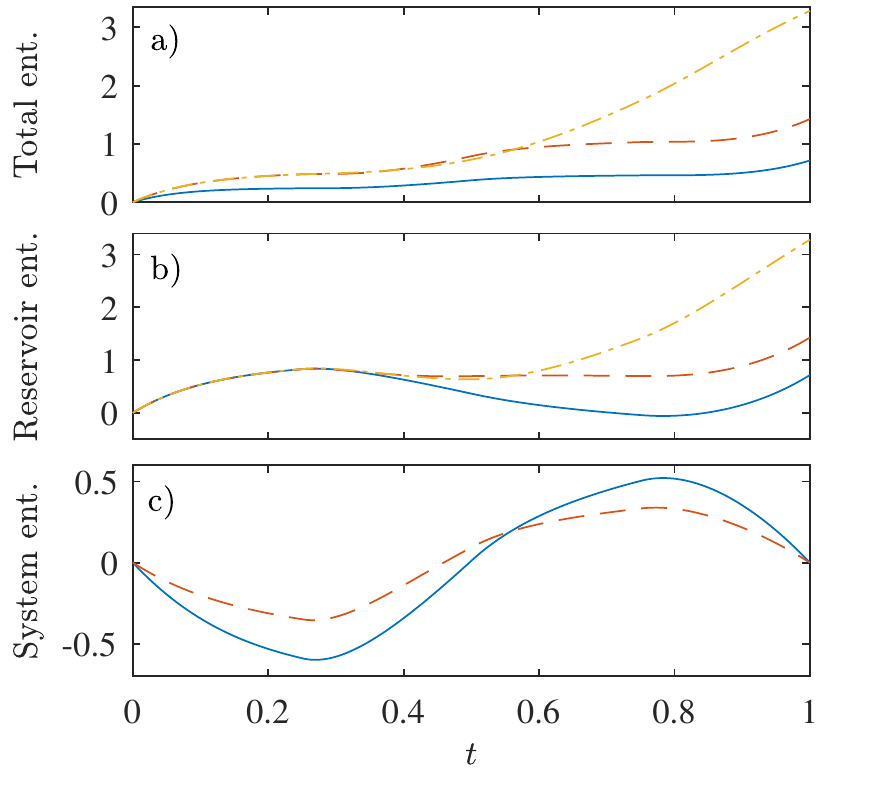}
\caption{Evolution of the various entropies discussed in the text as functions of  
time during the limit cycle, depicted in Fig.~\ref{fig:driving}, with $v_>=4$. Panel a) ``total'' ABE entropy changes $\Delta S^+_{\rm tot}$ for ``sailboats'', (red dashed line), $\Delta S^-_{\rm tot}(t)$ for ``surfboards'' (yellow dot-dashed line), both from Sec.~\ref{sec:surfsandsails}, and the effective entropy change $\Delta S^{\rm eff}_{\rm tot}(t)$ from Eq.~\eqref{eq:Stoteff} (solid blue line).
Panel b) shows corresponding changes in the reservoir entropy $\Delta S^+_{\rm R}(t)$ from Eq.~\eqref{eq:Sreven} (red dashed line), $\Delta S^-_{\rm R}(t)$ from Eq.~\eqref{eq:Srodd} (yellow dot-dashed line), and $\Delta S^{\rm eff}_{\rm R}(t)$ from integrating Eq.~\eqref{eq:SRGauss} (solid blue line), and panel c) in the system entropy $\Delta S^{\rm eff}$(t) from Eq.~\eqref{eq:DeltaSGauss} (solid blue line) and $\Delta S(t)$ from Eq.~\eqref{eq:DeltaS} (red dashed line).}
\label{fig:entropy_within_cycle}
\end{center}
\end{figure}

\begin{figure}[t]
\begin{center}
\centering
\includegraphics[width=1.0\columnwidth]{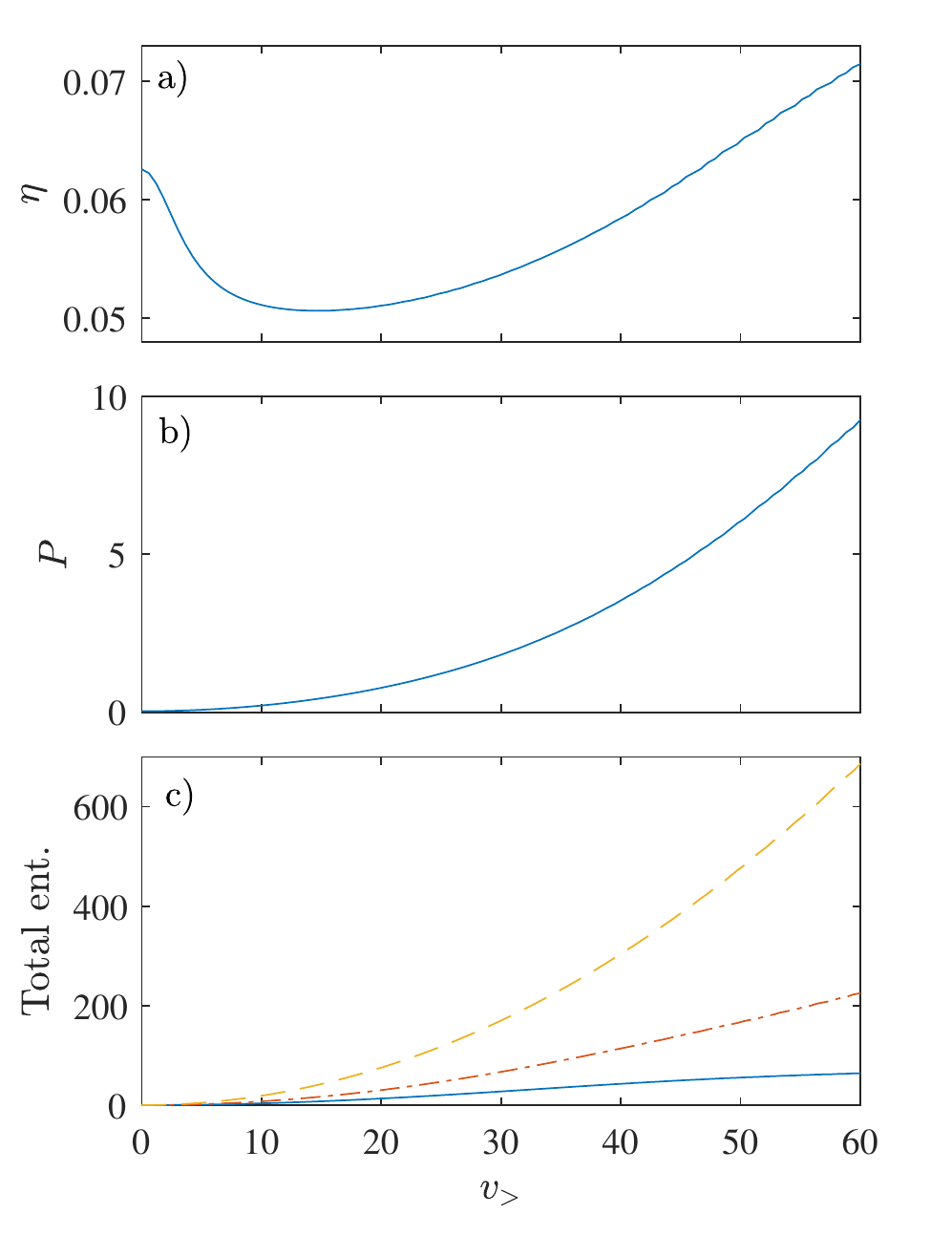}
\caption{Efficiency a), power output b), both from Eq.~\eqref{eq:PandEta}, and total entropy production c), as functions of the maximum active velocity $v_>$ for the protocol from Fig.~\ref{fig:driving}. Panel c) $\Delta S^-_{\rm tot}$ (yellow dashed), $\Delta S^+_{\rm tot}$ (red dot-dashed), and $\Delta S^{\rm eff}_{\rm tot}$ (solid blue), all from Eq.~\eqref{eq:integral}.}
\label{fig:performance_v}
\end{center}
\end{figure}

Let us finally investigate the most complex case of non-quasi-static cycles for which the protocol from Sec.~\ref{sec:driving} is imposed with cycle durations $t_{\text p}$ significantly shorter than the internal relaxation times $\tau_H = 1/(\mu k + D_{\rm r})$ and $\tau_\sigma = 1/(2\mu k)$ for $T_\text{eff}$ and $\sigma$, respectively. The ABE model provides full control over the finite-time thermodynamics.
To check our analytical results for the variance given in Sec.~\ref{sec:variance}, we compared it to direct numerical solutions of the equations of motion via the matrix numerical method of Ref.~\cite{hol18}, and found perfect agreement. We also note that the new features observed in the analytical results for the toy model are generic, and should qualitatively also be observed for other heat engines in contact with non-equilibrium reservoirs.

The hallmark of non-quasi-static operation of any thermodynamic heat engine is the observation of a net entropy increase during the cycle. Therefore,  Fig.~\ref{fig:entropy_within_cycle} depicts the individual entropy changes defined in Secs.~\ref{sec:equilirbium_model} and \ref{sec:entropyproduction} as functions of time during the limit cycle. Panel a) shows that both the total effective  entropy change $\Delta S_{\rm tot}^{\rm eff}(t)$, measured by the ABE user, and the total ABE entropy changes $\Delta S_{\rm tot}^{\pm}(t)$, corresponding to the sailboat and surfboard interpretations, are non-decreasing functions of time. They thus meet the expectation for valid total entropies according to the second law of thermodynamics. It is noteworthy, that the ABE entropy changes $\Delta S_{\rm tot}^{\pm}(t)$ are larger than the effective entropy change $\Delta S_{\rm tot}^{\rm eff}(t)$, at all times, even during the first part of the cycle, given by $t \in (0, 0.25)$, where the active velocity $v$ vanishes. 

As gleaned from the panel b), the rates of entropy change in the bath, with $\dot{S}^{\rm eff}_{\rm R}$ given by Eq.~\eqref{eq:SRGauss} and $\dot S^{\pm}_{\rm R}(t)$ given by Eqs.~\eqref{eq:Sreven} and \eqref{eq:Srodd}, are in that case all equal.
The inequality $\Delta S_{\rm tot}^{\rm eff}(t) < \Delta S^{\pm}_{\rm tot}(t)$ is then solely caused by the different changes of the system entropy
\begin{eqnarray}
\Delta S^{\rm eff}(t) &=& S^{\rm eff}(t) - S^{\rm eff}(0) = \log{\frac{\sigma(t)}{\sigma(0)}},
\label{eq:DeltaSGauss}\\
\Delta S(t) &=& S(t) - S(0),
\label{eq:DeltaS}
\end{eqnarray}
shown in the panel c), with $S^{\rm eff}(t)$ and $S(t)$ given by Eqs.~\eqref{eq:effentSys} and \eqref{eq:sys_ent}, respectively. 
For the remaining time [$t\in (0.25, 1)$] of the cycle, even the changes in the bath entropies $\Delta S^{\pm}_{\rm R}(t)$ of the ABE are larger than $\Delta S_{\rm R}^{\rm eff}(t)$. While $\dot{S}^{-}_{\rm R}(t) \ge \dot{S}_{\rm R}^{\rm eff}(t)$ and $\dot{S}_{\rm tot}^{\pm} \ge \dot{S}_{\rm tot}^{\rm eff}$ always hold, we find that $\dot{S}^{+}_{\rm R}(t) < \dot{S}_{\rm R}^{\rm eff}(t)$ is not ruled out (detailed data now shown). The figure also corroborates the periodicity of the system entropies $S^{\rm eff}(t)$ and $S(t)$, so that the total entropy changes $\Delta S^{\rm eff}_{\rm tot}(t_{\rm p})$ and $\Delta S^{\pm}_{\rm tot}(t_{\rm p})$ per cycle are solely determined by the (per cycle) entropy changes $\Delta S^{\rm eff}_{\rm R}(t_{\rm p})$ and $\Delta S^{\pm}_{\rm R}(t_{\rm p})$ in the bath, as it should be.

\begin{figure}[t]
\begin{center}
\centering
\includegraphics[width=1.0\columnwidth]{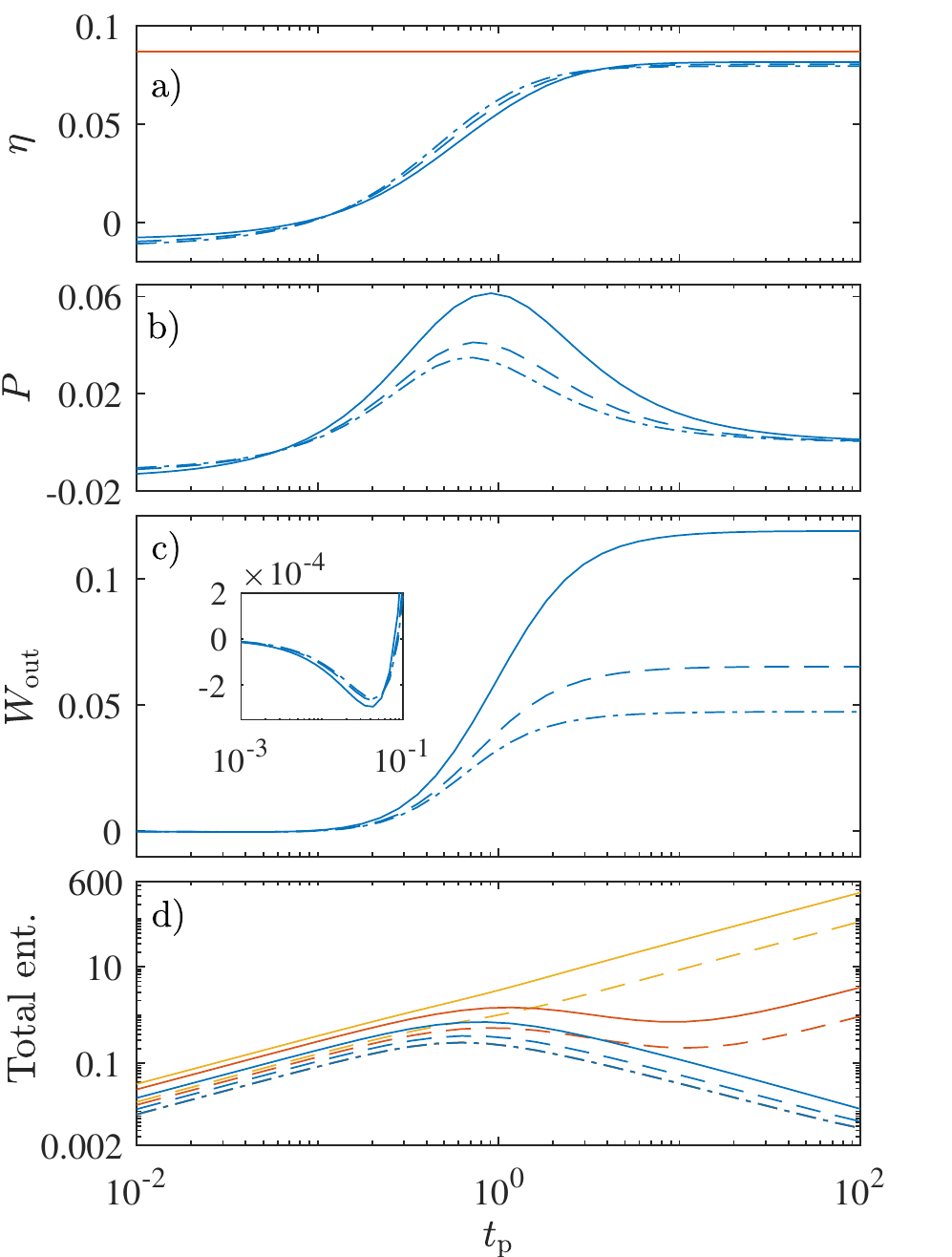}
\caption{Efficiency a), output power b), and output work c), defined in Sec.~\ref{sec:energetics}, and total entropy production d) for $v_>= 0$ (dot-dashed lines), $v_>= 2$ (dashed lines) and $v_>=4$ (solid lines) as functions of cycle duration $t_{\text p}$. The inset in panel c) magnifies the initial part of the plot for $t_{\rm p} \in [10^{-3},10^{-1}]$. Panel d) $\Delta S_{\text{tot}}^-$ (yellow), $\Delta S_{\text{tot}}^+$ (red), and $\Delta S_{\text{tot}}^{\text{eff}}$ (blue); all according to Eq.~\eqref{eq:integral}. Other parameters as in Fig.~\ref{fig:driving}.}
\label{fig:performance_tp}
\end{center}
\end{figure}

To study the influence of activity on the ABE performance, in Fig.~\ref{fig:performance_v}, we fix all the parameters according to Fig.~\ref{fig:driving} and vary the maximum active velocity $v_>$. For small values of $v_>$ the efficiency is decreased by the activity, while for large values of $v_>$ it is increased, and eventually attains a constant maximum value.
This behavior can be understood as follows. The efficiency of the heat engine quite generally increases with the largest difference in the effective temperature $\max{(T_{\rm eff})} - \min{(T_{\rm eff})}$, similarly as in the Carnot formula. Even beyond the quasi-static regime one expects that the effective temperature is qualitatively described by Eq.~\eqref{eq:TeffInf}. For small values of $v_>$, Eq.~\eqref{eq:TeffInf} implies that the temperature difference can be decreased by variations of the rotational diffusion coefficient, depicted in Fig.~\ref{fig:driving}c), while it increases with $v_>$ for large $v_>$. More intuitive behavior is observed for the power Fig.~\ref{fig:performance_v}b) and the entropy productions $\Delta S_{\text{tot}}^{\text{eff}}$ and $\Delta S_{\text{tot}}^{\pm}$ Fig.~\ref{fig:performance_v}c) that monotonically increase with $v_>$.

Finally, we assess the effect of the finite-time driving on the ABE operation. Specifically, in Fig.~\ref{fig:performance_tp}, we depict performance of the ABE as function of the cycle duration $t_{\rm p}$ for three values of the maximum active velocity $v_>$. In panel a), the efficiency monotonously increases with increasing $t_{\rm p}$ and eventually reaches the quasi-static limit (the red line). Notably, whether the efficiency is increased or decreased by the bath activity depends on the cycle duration, as evidenced by the dashed and solid lines wandering above and below the dot-dashed line.
Namely, apart from enhancing the output work and power [panels b) and c)], the activity also provides an increased heat flow into the system. As expected, the output power vanishes for large cycle durations and exhibits a maximum for a certain value of $t_{\rm p}$. On the contrary, the output work is, for large cycle times, an increasing function which converges to the quasi-static value, which monotonously increases with $v_>$. Interestingly, for $10^{-2}\lesssim t_{\rm p}\lesssim 10^{-1}$, the output work exhibits a shallow negative excursion as revealed by the blowup in the inset. This implies a lower bound $t_{\rm p}\approx 10^{-1}$ on the cycle duration, below which the system ceases to operate as a heat engine.

As can be observed in Fig.~\ref{fig:performance_tp}d), for small and large cycle durations, the cycle-time dependence of the total entropy productions $\Delta S_{\text{tot}}^{\text{eff}}(t_{\rm p})$ and $\Delta S_{\text{tot}}^{\pm}(t_{\rm p})$ exhibits asymptotic power-law behavior. Taylor expansions  of the total entropy productions in $t_{\rm p}$ and $1/t_{\rm p}$, respectively, give $\Delta S_{\text{tot}}^{\pm} \propto \Delta S_{\text{tot}}^{\text{eff}} \propto t_{\rm p}$ for short $t_{\rm p}$ and $\Delta S_{\text{tot}}^{\pm} \propto t_{\rm p}$ for $v \neq 0$, and $\Delta S_{\text{tot}}^{\text{eff}} \propto 1/t_{\rm p}$ regardless of $v$, for long $t_{\rm p}$. To be more specific, all the total entropy productions in question assume the form
\begin{equation}
\Delta S_{\rm tot}^{\rm z} = - \int_0^{t_{\rm p}} dt \frac{1}{\mathcal{T}(t)}\left(\dot{Q} + \mathcal{F}  \right)(t),
\label{eq:integral}
\end{equation}
where $\mathcal{T} = T_{\rm eff}$ and $\mathcal{F} = 0$ for ${\rm z} = {\rm eff}$,  $\mathcal{T} = T$ and $\mathcal{F} = - 2 \mu k (T_{\rm eff}- T)$ for ${\rm z} = -$, and $\mathcal{T} = T$ and $\mathcal{F} = 2 \mu k (T_{\rm eff}- T)-v^2/\mu$ for ${\rm z} = +$. For fast driving of the engine, ($t_{\rm p}$ much smaller than the intrinsic relaxation times), the colloid cannot react to the changing driving and settles on a time-independent state corresponding to a mean value of the driving. Hence, Eq.~\eqref{eq:integral} can be approximated for all ${\rm z}$ by $\Delta S_{\rm tot}^{\rm z} \approx - t_{\rm p}\left(\dot{Q} + \mathcal{F}  \right)/\mathcal{T}$, where the integrand is evaluated using the time-independent state attained for $t_{\rm p}\to 0$.

For slow driving ($t_{\rm p}$ much larger than the intrinsic relaxation times), the colloid attains its steady state~\eqref{eq:VarInf} independent of the cycle duration $t_{\rm p}$. Substituting the integration time $t$ in Eq.~\eqref{eq:integral} by the dimensionless time $\tau = t/t_{\rm p}$ yields
\begin{equation}
\Delta S_{\rm tot}^{\rm z} = - t_{\rm p} \int_0^1 d\tau \frac{1}{\tilde{\mathcal{T}}(\tau)}\left(\frac{1}{t_{\rm p}}\frac{d\tilde{Q}(\tau)}{d\tau} + \mathcal{\tilde{F}}(\tau)  \right),
\label{eq:integral2}
\end{equation}
where $\tilde{\mathcal{T}}(\tau) = \mathcal{T}(\tau t_{\rm p})$, $\tilde{Q}(\tau) = Q(\tau t_{\rm p})$, and $\tilde{\mathcal{F}}(\tau) = \mathcal{F}(\tau t_{\rm p})$. The effective total entropy production $\Delta S_{\rm tot}^{\rm eff}$ vanishes in the limit $t_{\rm p} \to \infty$, and thus the leading contribution in Eq.~\eqref{eq:integral2} is expected to be of order $1/t_{\rm p}$. Indeed, expanding under the integral, we obtain (for $\mathcal{F} = 0$) 
\begin{equation}
\frac{1}{\tilde{\mathcal{T}}}\frac{d\tilde{Q}}{d\tau}
\approx \frac{d}{d\tau} \log \sigma_{\infty}
+ \frac{1}{t_{\rm p}} C.
\label{eq:expansion}
\end{equation}
Since the first term represents a total derivative, the corresponding loop integral vanishes and what remains is the correction $C/t_{\rm p}$ with a $t_{\rm p}$ independent constant $C$. For ${\rm z} = \pm$, the leading contribution to the integral~\eqref{eq:integral2}
is simply determined by the non-zero value of $\lim_{t_{\text p} \to \infty}\mathcal{F}$ and thus we find $\Delta S_{\rm tot}^{\pm} \propto t_{\rm p}$ for large $t_{\rm p}$. For $v = 0$, all three definitions of entropy production are equivalent since then $\mathcal{T} = T=T_{\text{eff}}$ and $\mathcal{F}=0$. This proves the scalings found in Fig.~\ref{fig:performance_tp}d).

\section{Conclusion and outlook}
\label{sec:conclusion}

\textcolor{black}{
We argued on a very general basis that energy extracted from non-equilibrium reservoirs by cyclically operating engines qualifies as heat only if there exists a precise mapping to an equivalent cycle with an equilibrium bath at a time-dependent effective temperature, which yields the same power and efficiency. We have discussed the most general setting when such a mapping always exists and explained that engines which do not allow for a consistent definition of effective temperature should rather be understood as (possibly loss-making) work-to-work converters than heat engines. A benefit of the effective-temperature mapping is that conventional bounds on both the finite-time and the quasi-static thermodynamic performance of machines, especially heat engines, become applicable to those with non-equilibrium (active) baths \cite{kri16,zak17,mar18,Kumari2019}. As a part of our discussion, we have therefore been able to provide a new perspective on recent claims of surprisingly high Stirling efficiencies (surpassing the second law bound corresponding to infinite temperature steps) in a bacterial heat engine that was experimentally realized by Krishnamurthy et al.~\cite{kri16}.}

To exemplify the general findings, we have derived a simple strategy to map the average thermodynamics of a linear Langevin system with arbitrary additive noise to an effective equilibrium system. The mapping is based on the matching of the dynamical equations for the second moment of position, which happens to determine the (average) energetics. It is valid for arbitrary protocols imposed by the time-dependent model parameters. In the quasi-static limit, the (generally time-dependent) effective temperature $T_{\rm eff}(t)$ \eqref{eq:Teffgen} that accomplishes the mapping recovers the known expression \eqref{eq:qs_Teff}. 

We have further exemplified these somewhat abstract general notions by a fully worked example
of a specific engine design that we call
the ABE, since the particle dynamics is based on the well-known active Brownian particle (ABP) model. Our qualitative conclusions should carry over to other designs, though. In particular, we find that the explicitly computed effective temperature $T_{\rm eff}$ has some non-intuitive features. (i) During the limit cycle, which is attained by the ABE at long times, it obeys a first-order differential equation and thus acquires some time dependence $T_{\rm eff}(t)$ with a technically relevant characteristic relaxation time. (ii) It is important to realize that it can therefore vary in time even during those parts of the cycle in which the model parameters are held constant. (iii) Even in the quasi-static limit, $T_{\rm eff}$ depends on the strength of the potential. This means that realizing specific thermodynamic conditions, like an ``isothermal'' process with respect to the effective temperature, is generally not trivial.

The ABE model is also instructive with respect to some limitations of the effective-bath mapping. Namely, by construction, the latter is blind to the potentially rich features of the non-equilibrium bath beyond the second moment of the particle position, which we identified as the working degree of freedom of the engine. The effective description thus misses the non-Gaussian shape of the positional probability density and the corresponding Shannon entropy, for example, and also all housekeeping heat fluxes required to maintain the bath activity. Accordingly, we could demonstrate that the entropy production in the effective model can be understood as a lower bound for all conceivable practical and theoretical realizations. Namely, it vanishes upon quasi-static operation, whereas any detailed model of the bath dynamics would, like the explicitly studied ABE, necessarily reveal some of the housekeeping heat fluxes and their associated entropy production.

\textcolor{black}{
As an outlook, it would be interesting to study possible generalizations of our analysis of the linear model for arbitrary time-dependent friction kernels and correlation matrices, thus also including under-damped dynamics, which does not belong into the class of systems where the effective temperature always exists. Another possible extension could be the application of the presented method to non-linear systems, e.g., by deriving approximate time-dependent effective temperatures via suitable closures of the equations describing the relevant degrees of freedom. Our general analysis shows that at least for Hamiltonians of the form $\mathcal{H} = k(t) h(\mathbf{x})$, with an arbitrary function $h(\mathbf{x})$ diverging at $|x| \to \infty$, this should always be possible.}

\begin{acknowledgments}
We acknowledge funding by Deutsche Forschungsgemeinschaft ({DFG}) via SPP 1726/1 and KR 3381/6-1, and by Czech Science Foundation (project No. 20-02955J). VH gratefully acknowledges support by the Humboldt foundation. S.S. acknowledges funding by International Max Planck Research Schools ({IMPRS}).
\end{acknowledgments}

\bibliography{HE_ref}


\appendix
\section{Analytical solution for variance}
\label{appx:var_der}

Inserting the time correlation matrix~\eqref{eq:2tcorr} for the ABP model into Eq.~\eqref{eq:r_formal2}, Eq.~\eqref{eq:var_formal}
yields the following dynamic equation for the variance $\sigma =\left\langle \mathbf{r}\cdot\mathbf{r}\right\rangle = \left\langle x^2+y^2\right\rangle$: 
\begin{multline}
\dot{\sigma} + 2\mu k \sigma = 4 \left<x_0 \eta_x(t)+ y_0 \eta_y(t) \right>{\rm e}^{-K(t,t_0)} + 4D(t) \\ 
+ 2v(t)\int_{t_0}^{t} dt' v(t'){\rm e}^{- F(t,t')}\,,
\label{eq:dotsigma}
\end{multline}
where
\begin{eqnarray}
K(t,t_0) &=& \mu \int_{t_0}^{t} dt' k(t')\,,
\label{eq:K}\\
F(t,t_0) &=& K(t,t_0) + \int_{t_0}^{t} dt' D_{\text{r}}(t')\,.
\label{eq:F}
\end{eqnarray}

In order to explicitly evaluate the thermodynamics of the particular realization of an active Brownian heat engine described in Sec.~\ref{sec:example}, namely the ABP-based engine that we refer to as the ABE model, we need the solution of Eq.~\eqref{eq:dotsigma}. More precisely, we can concentrate onto the time periodic solution, which is attained by the system at late times, after transients have relaxed, so that it settles onto a limit cycle (c.f. Fig.~\ref{fig:var}). Taking the limit $t_0 \to -\infty$ in the formal solution to Eq.~\eqref{eq:dotsigma}, we obtain
\begin{equation}
\sigma(t) = 2\lim_{t_0 \to - \infty}\int_{t_0}^{t} dt' [2 D(t') + v(t') H(t')] {\rm e}^{-2K(t,t')}
\label{eq:var_exact1}
\end{equation}
with
\begin{equation}
H(t) = \lim_{t_0 \to - \infty}\int_{t_0}^{t} dt' v(t'){\rm e}^{- F(t,t')}\,.
\label{eq:H}
\end{equation}
For the numerical evaluation of Eq.~(\ref{eq:var_exact1}) it is useful to exploit that $H(t)$ is a $t_{\rm p}$-periodic function and to rewrite $K(t,t_0)$ as $K(t,t_0) = {\lfloor}(t - t_0)/t_{\rm p}{\rfloor} K(t_{\rm p}, 0) + K(t,t_0 + {\lfloor}(t - t_0)/t_{\rm p}{\rfloor} t_{\rm p})$ using the $t_{\rm p}$-periodicity of $k(t)$ (the symbol ${\lfloor}x{\rfloor}$ denotes the floor operation) and similarly for $F(t,t_0)$. 
Interestingly, using a simple trick, the time-periodic late-time limit can be found without considering the (numerically inconvenient) limit $t_0 \to -\infty$, just as in the case of memoryless dynamics \cite{sch08,hol14}. \textcolor{black}{The key insight is that, in the long-time regime, the functions $\sigma$ and $H$ obey two coupled ordinary differential equations, namely
Eqs.~\eqref{eq:dynH} and \eqref{eq:dyn} in Sec.~\ref{sec:variance}, which follow from Eqs.~\eqref{eq:var_exact1} and \eqref{eq:H} by taking derivative with respect to $t$.}


\section{Slow driving limit of variance}
\label{appx:var_approx}

For slowly varying driving functions $k(t)$, $D(t)$, $D_{\text{r}}(t)$ and $v(t)$, the variance (\ref{eq:var_exact1}) can be approximated using a simple formula which follows from the Laplace type approximation of the integral \cite{ble75, nem12}
\begin{multline}
\int_{t_0}^t dt\, f(t'){\rm e}^{\int_{t'}^tdt''\, g(t'')} = \int_{t_0}^t dt\, f(t'){\rm e}^{t_{\rm p} \int_{t'/t_{\rm p}}^{t/t_{\rm p}}dt''\, g(t_{\rm p} t'')} =
\\
\frac{f(t)}{g(t)} - \frac{1}{g^2(t)}
\left[ \dot{f}(t) - f(t) \frac{\dot{g}(t)}{g(t)} \right]
+ o(\dot{f}, \dot{g}).
\label{eq:approx_int}
\end{multline}
Applying this approximation first on the function $H(t)$ (\ref{eq:H}) and then on the variance $\sigma(t)$ (\ref{eq:var_exact1}), we obtain the approximate result 
\begin{multline}
\sigma(t) = \sigma_{\infty} - \frac{v^2}{k\mu\kappa^2}\left(\frac{\dot{v}}{v} -\frac{\dot{\kappa}}{\kappa}\right) - \frac{D}{k^2\mu^2}\left(\frac{\dot{D}}{D} -\frac{\dot{k}}{k}\right) \\
-\frac{v^2}{2 k^2\mu^2\kappa}\left(2\frac{\dot{v}}{v} - \frac{\dot{\kappa}}{\kappa} - \frac{\dot{k}}{k} \right) + o(\dot{v},\dot{D},\dot{k},\dot{\kappa}).
\label{eq:1stordervar}
\end{multline}
Here, $\sigma_{\infty}$ is the variance~\eqref{eq:VarInf} for infinitely slow driving and $\kappa = \kappa(t) = k\mu+ D_{\text{r}}$. For discontinuous driving, the limiting solution $\sigma_{\infty}$ is also discontinuous. The first order correction (\ref{eq:1stordervar}) may also be discontinuous if the first derivatives of the driving functions exhibit jumps. In such a case, however, the assumption on the smallness of the derivatives used in the calculation leading to Eq.~(\ref{eq:1stordervar}) is not valid. In accord with the discussion below Eq.~(\ref{eq:FP}) in App.~\ref{appx:PDF}, Eq.~(\ref{eq:1stordervar}) reveals that activity-corrections are at least second order in $v$. 

\section{Entropy production from path probabilities}
\label{appx:entropy}

The entropy 
\begin{equation}
\Delta S_{\rm R, \Gamma}(t) = \log (P_{\rm F}/P_{\rm R}) 
\label{eq:SRappx}
\end{equation}
 delivered to the bath by a particle moving along a trajectory $\Gamma(t) = \{\textbf{r}(t'),\theta(t')\}_{t'=0}^t$ of the stochastic process \eqref{eq:Langevin1A}, \eqref{eq:Langevin1B} is given by the logarithm of the ratio of conditional probabilities $P_{\rm F}$ and $P_{\rm R}$~\cite{mae03,lup13}, for the trajectory conditioned with respect to its initial point and its time-reversed image. 
 Up to normalization, the forward probability is given by
\begin{equation}
P_{\rm F} \propto  {\rm e}^{-2\int_0^t dt'\, \left[\bm{\xi}\cdot\bm{\xi} + \xi_\theta^2\right]},
\end{equation}
where the noise terms $\bm{\xi} = [\dot{\textbf{r}} + \mu \nabla_{\textbf r} {\mathcal H} - \textbf{v}]/\sqrt{2D}$ and $\xi_\theta = \dot{\theta}/\sqrt{2 D_{\rm r}}$ follow from Eqs.~\eqref{eq:Langevin1A} and \eqref{eq:Langevin1B} \cite{kle95}. The backward probability is given by a similar formula. One just has to change the sign before quantities which are odd with respect to time reversal.

Assuming the active velocity $\textbf{v} = v (\cos\theta, \sin \theta)$ to be time-reversal even, the odd variables in Eqs.~\eqref{eq:Langevin1A} and \eqref{eq:Langevin1B} are only time derivatives, giving
\begin{equation}
\left(P_{\rm F}/P_{\rm R}\right)^+ = {\rm e}^{-\int_0^t dt'\, \left(\nabla_{\textbf r} {\mathcal H} - \textbf{v}/\mu\right) \cdot \dot{\textbf{r}}/T},
\label{eq:ratioVeven}
\end{equation}
whereas, for time-reversal odd $\mathbf{v}$, we find 
\begin{equation}
\left(P_{\rm F}/P_{\rm R}\right)^- = {\rm e}^{-\int_0^t dt'\, \nabla_{\textbf r} {\mathcal H} \cdot\left(\dot{\textbf{r}} - \textbf{v}\right)/T}.
\label{eq:ratioVodd}
\end{equation}
The entropy delivered to the reservoir during time interval $(0,t)$ follows as
\begin{equation}
\Delta S_{\rm R}(t) = \left< \Delta S_{\rm R, \Gamma}(t)\right>_{\rm \Gamma} = 
\left<\log (P_{\rm F}/P_{\rm R}) \right>_{\rm \Gamma},
\label{eq:sRapp}
\end{equation}
where the average is taken over the individual realizations $\Gamma$ of the stochastic process \cite{lup13}. With Eq.~\eqref{eq:ratioVeven} for the time-even active velocity, it yields
\begin{equation}
\Delta  S_{\rm R}^{+}(t) = \int_0^t dt'\, \frac{1}{T} \left< \left(\frac{\textbf{v}}{\mu} - \nabla_{\textbf r} {\mathcal H}\right) \cdot \dot{\textbf{r}} \right>,
\label{eq:S+app}
\end{equation}
and with Eq.~\eqref{eq:ratioVodd}, for the time-odd active velocity,
\begin{equation}
\Delta  S_{\rm R}^{-}(t) = \int_0^t dt'\, \frac{1}{T}\left<   (\dot{\textbf{r}} - \textbf{v}) \cdot (- \nabla_{\textbf r} {\mathcal H})\right>\;.
\label{eq:S-app}
\end{equation}

\section{Probability distributions (PDFs)}
\label{appx:PDF}

\begin{figure}[t]
\begin{center}
\centering
\includegraphics[width=1.0\columnwidth]{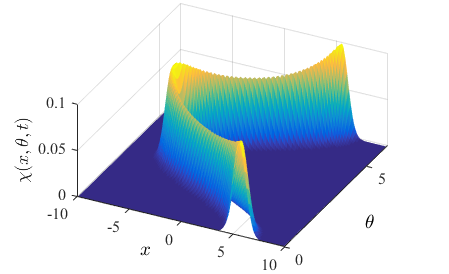}
\caption{Probability distribution $\chi$ for particle position $x$ and orientation $\theta$ at the end of the hot isotherm ($t=3t_{\rm p}/4$, see Fig.~\ref{fig:driving}). We take $v_>=30$ and $t_{\rm p} = 10^4$. Other parameters are the same as in Fig.~\ref{fig:driving}.}
\label{fig:distriburion3D}
\end{center}
\end{figure}

\begin{figure}[t]
\begin{center}
\centering
\includegraphics[width=1.0\columnwidth]{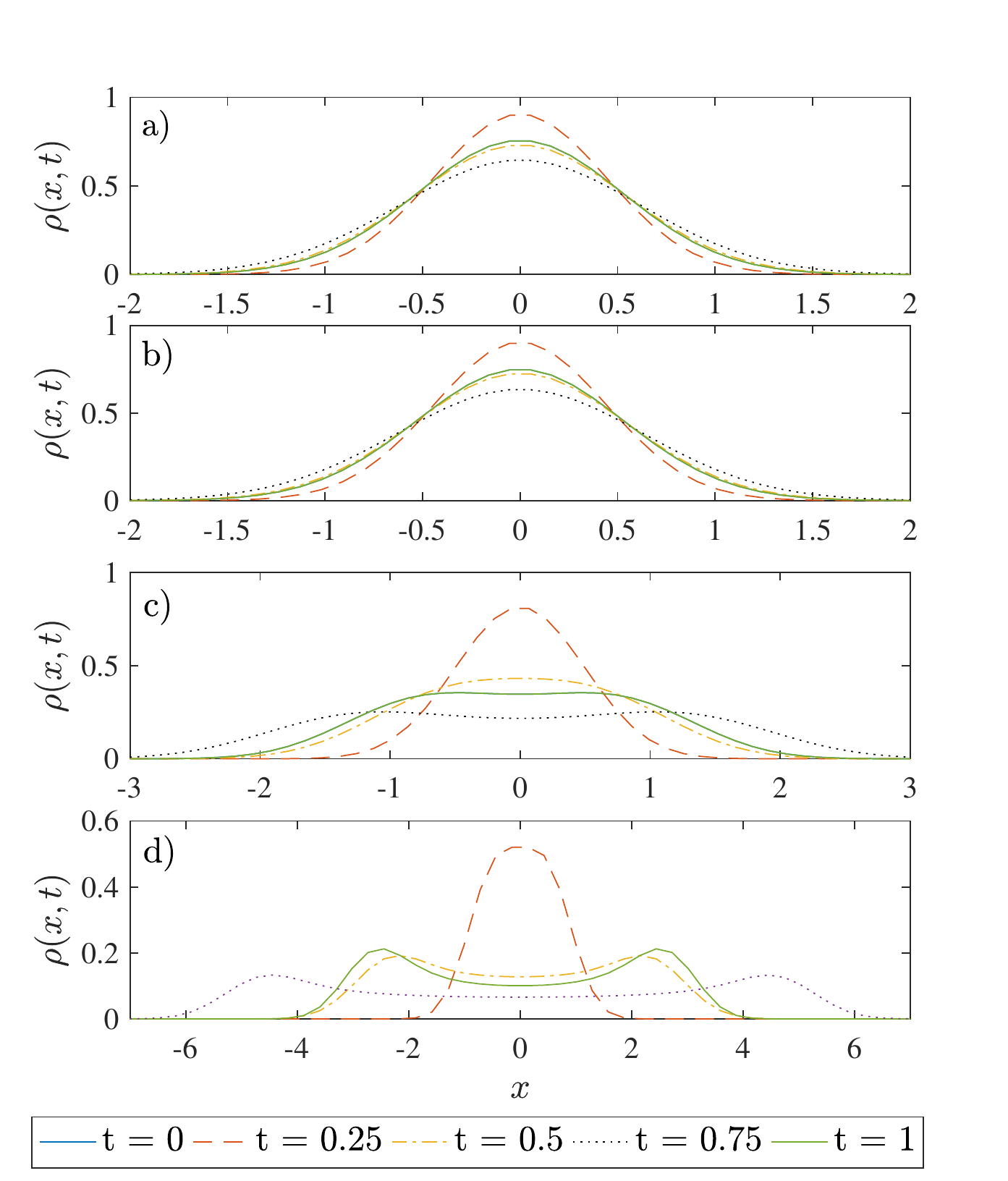}
\caption{Marginal distribution $\rho$ for the particle position $x$ at the end of the individual branches of the cycle for different values of the maximum active velocity a) $v_>=0$, b) $v_>=1$, c) $v_>=10$, and d) $v_>=30$. We have set $t_{\rm p} = 1$, corresponding to non-stationary driving, other parameters as in Fig.~\ref{fig:driving}. Note that the curves at $t=0$ and $t=1$ are equal, in accord with the time periodic operation.}
\label{fig:distriburion}
\end{center}
\end{figure}

In the 3-dimensional Langevin system \eqref{eq:Langevin1A}--\eqref{eq:Langevin1B}, the $x-y$ coordinates  are coupled via the active velocity $\textbf{v}$. The steady probability distribution (PDF) to find the particle with orientation $\theta$ at position $(x,y)$ thus cannot generally be written in the separated form $p(x,y,\theta) = \chi(x,\theta)\iota(y,\theta) = \chi(x,\theta) \chi(y,\pi/2 - \theta)$, where $\chi(x,\theta)$ solves the 2-dimensional Fokker--Planck equation 
\begin{equation}
\partial_t \chi = \left[D\partial^2_{x} + D_{\text{r}}\partial^2_{\theta} + \partial_x\left(\mu k \partial_x x - v \cos\theta \right)
\right] \chi \;.
\label{eq:FP}
\end{equation}
Inserting the separation ansatz into the 3-dimensional equation \eqref{eq:FPE} and using the formula~\eqref{eq:FP} leads to the condition $2D_{\rm r} \partial_\theta\chi(x,\theta)\partial_\theta \iota(y,\theta) = 0$ that cannot be fulfilled in general. Nevertheless, one can still reduce the 3-dimensional system to just two degrees of freedom by introducing the polar coordinates $x=r\cos\phi,\,y=\sin\phi$. Then, Eq.~\eqref{eq:Langevin1A} transforms to
\begin{align}
\dot{r}&=-\mu kr+v\cos(\theta-\phi)+\sqrt{2D}\eta_r\,,
\label{eq:pola}\\
\dot{\phi}&=\frac{v}{r}\sin(\theta-\phi)+\sqrt{\frac{2D}{r^2}}\eta_\phi\,,
\label{eq:polb}
\end{align}
while $\theta$ still obeys Eq.~\eqref{eq:Langevin1B}. The symbols $\eta_r$ and $\eta_\phi$ denote independent, zero-mean, Gaussian white noises. Since Eqs.~\eqref{eq:pola} and \eqref{eq:polb} only depend on the difference $\theta-\phi$, introducing the relative angle $\psi=\theta-\phi$, subject to the  zero-mean, Gaussian white noise $\eta_\psi$ renders them in the form
\begin{align}
\dot{r}&=-\mu k r+v\cos\psi+\sqrt{2D}\eta_r\,,\\
\dot{\psi}&=-\frac{v}{r}\sin\psi+\sqrt{2\left(\frac{D}{r^2}+D_{\rm r}\right)}\eta_\psi\,.
\label{eq:pol2}
\end{align}
The corresponding Fokker--Planck equation for the PDF $\rho=\rho(r,\psi,t)$ reads \cite{ris96}
\begin{multline}
\partial_t\rho=\left[D\partial^2_r+\left(\frac{D}{r^2}+D_{\rm r}\right)\partial^2_\psi\right]\rho-\cos\psi\partial_r(v\rho)\\-D\partial_r\left(\frac{\rho}{r}\right)+\mu k\partial_r(r\rho)+\frac{v}{r}\partial_\psi(\sin\psi\rho)\,.
\label{eq:FPEpol}
\end{multline}
In general, the equations \eqref{eq:FP} and \eqref{eq:FPEpol} [or, equivalently \eqref{eq:FPE}] can not be solved analytically and thus we solved them using the numerical method described in Ref.~\cite{hol18}. We compared the numerical solution of Eq.~\eqref{eq:FPEpol} to the separated ansatz $p(x,y,\theta) = \chi(x,\theta) \chi(y,\pi/2 - \theta)$ and found out that, although not exact, the ansatz describes the full 3-dimensional PDF $p(x,y,\theta)$ sufficiently well. Since the 2-dimensional PDF allows for a more intuitive discussion and exhibits the main qualitative features of $p(x,y,\theta)$, we restrict the following discussion to $\chi(x,\theta)$.

Figure~\ref{fig:distriburion3D} shows a snapshot of the PDF $\chi(x,\theta,t)$, solution of \eqref{eq:FP}, at the end of the third branch of a quasi-static cycle introduced in Sec.~\ref{sec:driving} (the hot ``isotherm''). The figure reveals the typical shape of the PDF $\chi$, with two global maxima located at $\theta= 0$ and $\pi$, which survives even for rapid driving protocols. Physically, the shape of the PDF can be understood as follows: 1) for any fixed orientation angle $\theta$, the PDF can be expected to exhibit a maximum at the position where the active velocity (which acts in the Langevin Eq.~\eqref{eq:Langevin1A} for $x$ as a force $v \cos \theta/\mu$) is balanced by the force $kx$ exerted by the parabolic potential; 2) the projection $v \cos \theta/\mu$ of $\mathbf{v}$ on the $x-$coordinate changes slowest around its extrema (0 and $\pi$), and thus most trajectories contribute to the surroundings of these points, making the extrema for 0 and $\pi$ largest.

Figure~\ref{fig:distriburion} shows snapshots of the marginal PDF $\rho(x,t) = \int d\theta \chi(x,\theta,t)$ for the position $x$ at the beginning of the individual branches of the cycle, for four values of the maximum active velocity $v_>$. With increasing $v_>$, the resulting PDFs become increasingly non-Gaussian and finally even exhibit two separated peaks. Physically, this behavior can be understood by the wall accumulation effect due to the persistence of the active motion \cite{fil14, wen08, nos16}, which creates the double peak during the cycle branches with large $v_>$.  (For similar PDFs, see Ref.~\cite{Shin2017, zhe13}.) Qualitatively similar results are also obtained in the quasi-static limit, as already apparent from Fig.~\ref{fig:distriburion3D}.

To get some intuition about these results on analytical grounds, we now present several approximate solutions to Eq.~\eqref{eq:FP}. Different from the standard diffusion ($v=0$) in an external potential, the quasi-static ($\partial_t \chi=0$) solution of the Fokker-Planck equation \eqref{eq:FP} is not given by the Boltzmann PDF. This is because one cannot subsume the activity into a generalized potential $\tilde{{\mathcal H}}$ which would act as a Lyapunov functional for the dynamics of $x$ and $\theta$. Nevertheless, there are several limiting cases where the Boltzmann form $\chi \propto \exp(-\tilde{{\mathcal H}}/T)$ is still a useful approximation.

The best analytical insight into the described qualitative properties of the presented numerical solutions to Eq.~\eqref{eq:FP} with time-dependent parameters is obtained for rotational diffusion coefficient $D_{\text{r}}$ much smaller than $k\mu$, corresponding to the limit of large $\mathcal{K}$ in Eq.~\eqref{eq:Kcal}. Then, the direction of the active velocity can be treated as quenched, so that the activity can be subsumed into a generalized potential $\tilde{{\mathcal H}} = kx^2/2 - v x \cos \theta/\mu$. The corresponding quasi-static solution of Eq.~(\ref{eq:FP}) then reads
\begin{equation}
\chi = \frac{1}{Z_\chi} \exp\left(\frac{v x \cos \theta}{\mu T}-\frac{kx^2}{2T} \right)\;,
\label{eq:distEQL}
\end{equation}
with a normalization constant $Z_\chi$. For each fixed value of the angle $\theta$, the PDF is then Gaussian with its maximum value $\exp[v^2 \cos^2\theta/(2T\mu^2k)]/{Z_\chi}$ at the position $v \cos \theta/(\mu k)$. The PDF thus posses two global maxima located at $(x,\theta) = [v/(\mu k),0]$ and $(x,\theta) = [-v/(\mu k),\pi]$, and is qualitatively similar to the PDF shown in Fig.~\ref{fig:distriburion3D}.

The marginal PDF for $x$ obtained from \eqref{eq:distEQL} then reads
\begin{equation}
\rho(x,t) = \int d\theta \chi = \frac{1}{Z_\rho} \exp\left(-\frac{k x^2}{2T}\right) I_0\left(\frac{v x}{\mu T}\right).
\label{eq:distEQLApp}
\end{equation}
Here, $I_0(x)$ denotes the modified Bessel function of the first kind and $Z_\rho$ is another normalization constant. The marginal PDF is Gaussian for $v=0$, and becomes more and more non-Gaussian with increasing $v/(\mu k)$. For large values of $v/(\mu k)$, it can even become bimodal. This behavior can be traced back to the shift of the maxima of the PDF $\chi$ with increasing $v/(\mu k)$. For small $v/(\mu k)$, the two maxima substantially overlap and the integration over the angle $\theta$ yields a single peak which is nearly Gaussian. For large values of $v/(\mu k)$, the two peaks do not overlap any more and the marginal PDF thus also exhibits two peaks. The behavior of the marginal PDF obtained in the limit $D_{\text{r}}\ll \mu k$ thus shows qualitatively the same behavior as the solution of Eq.~\eqref{eq:FP} shown in Fig.~\ref{fig:distriburion}.

For $D_{\text{r}}$ much larger than $k\mu$, corresponding to the limit of small $\mathcal{K}$ in Eq.~\eqref{eq:Kcal}, the quasi-static PDF is given by $\chi \propto \exp(-{\mathcal H}/T_{\text{eff}})$. This is because the rotational diffusion obliterates any persistence of the active motion, and the non-equilibrium bath effectively behaves like an equilibrium one with the renormalized temperature $T_{\text{eff}} = T + v^2/(2\mu D_{\text{r}})$. In this limit, the degrees of freedom $x$ and $y$ also become independent.

Yet another case admitting an analytical solution of Eq.~(\ref{eq:FP}) is that of quasi-static driving at small active velocity. Then the quasi-static PDF $\rho$ can be approximated by the McLennan-type form $\chi \approx \exp(- {\mathcal H}/T)[1 - W(x)]$ \cite{McLennan1959,kom08,Maes2010,siv12, nak13}. Without going into details, the function $W(x)$ is in general proportional to the (average) dissipation in the driven system \cite{Maes2010}, which, in our case, is given by the product of the active ``force'' $\mu^{-1} v \cos \theta$ and the particle velocity $\dot{x}$. Since the average over the angle $\theta$ of the active force is zero, 
the correction $W(x)$ to the particle PDF is seen to be at least second order in $v$. 

\end{document}